\newcommand{\project}{Practical Quantum Topological Data Analysis with Applications to High-Dimensional Feature Extraction and Time Series Analysis}

\documentclass[11pt]{article}
\usepackage[utf8]{inputenc}

\usepackage[letterpaper, top=.25in,bottom=1in,right=1in,left=1in,headsep=1in,includehead]{geometry}

\usepackage{fancyhdr}
\usepackage{graphicx}
\usepackage{xcolor}
\usepackage{comment}
\usepackage{float}
\usepackage{booktabs}
\usepackage{amsmath}
\RequirePackage[colorlinks=true, allcolors=blue]{hyperref}
\usepackage{array}
\usepackage{multirow}
\usepackage{algorithm}
\usepackage{algpseudocode}
\usepackage{bm}
\usepackage{amsmath,amssymb}
\usepackage[toc,page]{appendix}

\usepackage[absolute]{textpos}
\usepackage{datetime}

\newdateformat{monthyeardate}{%
  \monthname[\THEMONTH] \THEYEAR}

\newcommand{\bra}[1]{\langle#1\rvert} 

\title{\project
\\\vskip 12pt\large{\reporttype}}
\date{\today}

\usepackage{listings}
\usepackage{color}
\usepackage{pagecolor}

\usepackage{pythonhighlight}

\definecolor{dkgreen}{rgb}{0,0.6,0}
\definecolor{gray}{rgb}{0.5,0.5,0.5}
\definecolor{mauve}{rgb}{0.58,0,0.82}

\definecolor{ionqorange}{HTML}{FF5000}
\definecolor{ionqbackground}{RGB}{19,25,31}

\newcommand{\ket}[1]{|#1\rangle}

\fancyheadoffset[R]{1.2cm}
\fancyfootoffset{1.4cm}

\begin{document}

\clearpage 
\color{black} 


\pagestyle{fancy}
\fancyhf{}
\cfoot{\thepage}

\begin{center}

{\LARGE \bf \project \par}
\vspace{4ex}
\large
Jason Iaconis,
Sayonee Ray,
Samwel Sekwao,
Claudio Girotto,
Martin Roetteler\\
\vspace{2ex}
\large
IonQ Inc, College Park, MD 20740, USA\\
\vspace{2ex}

\end{center}


\begin{abstract}
Topological data analysis (TDA) provides a powerful framework for extracting information about the shape of complex, unstructured data, but the classical cost of computing high-dimensional topological features limits its application. Quantum algorithms for TDA offer a route around this bottleneck, yet existing approaches typically focus on exact or high-precision Betti number estimation, making the regime for practical quantum advantage appear narrow. Here, we instead frame quantum TDA as a feature-extraction method for downstream data analysis by extracting low-order spectral information from the combinatorial Laplacian as a proxy for high-dimensional topology.

We support this perspective from both the application and algorithmic sides. First, we show that higher-order TDA features improve predictive performance in two time-series applications: functional MRI analysis for neurodegenerative disease classification and financial time-series analysis for identifying market instability. Second, we develop a moment-based quantum algorithm and show that low-order moments, including the relative trace, are strongly correlated with high-dimensional Betti information, even when the relative Betti number is small. Finally, we present circuit constructions, resource estimates, quantum-classical crossover projections, and experimental results from a Barium development system similar to the forthcoming IonQ Tempo line, extracting Laplacian-derived observables from graph instances and quantitatively comparing them with exact Betti information.

Together, these results establish quantum TDA as a practical approach for extracting topological features from classically challenging data.
\end{abstract}

\newpage
\setcounter{tocdepth}{3}
\tableofcontents
\newpage


\section{Introduction}

As the popularity of data-driven science has grown over the past decade, so has the desire for more advanced tools for extracting useful structure from complex data.  Machine learning and AI models have shown tremendous success in learning from data for many classes of problems, however there still exist critical domains where our current techniques struggle. One such domain is in the modeling of complex systems, where the large scale behavior of the whole depends on a complicated set of dependencies among many interacting components. These complex systems can often be represented as networks where nodes represent components and edges represent their interactions, or mapped to point clouds in high dimensional spaces where distance encodes similarity between data points.  In recent years, the community has begun to appreciate that higher order interactions beyond simple pairwise relationships may be critical for understanding the behavior of systems which can be represented by complex networks \cite{bick2023higher, sizemore2019importance, bhattacharya_2024}.

Topological data analysis (TDA) provides a natural framework for studying this kind of higher-order structure.  Rather than describing a data set only through individual points or pairwise relationships, TDA calculates features which depend on the arrangement of many points at once. These features, known as topological invariants, are computed using tools from algebraic topology and provide a quantitative description of the global shape of a graph, network, or point cloud. In this sense, TDA can reveal structure that is invisible to purely local or pairwise summaries, such as connected
components, loops, enclosed voids, and their higher-dimensional analogues. These features can then be used, for example, by downstream AI models to better predict or classify the behavior of the underlying complex system. These methods have been used successfully across a wide range of fields \cite{chazal2021introduction,skaf2022topological,lum2013extracting}. However, most practical applications have focused on the lowest-dimensional topological features, especially connected components and loops~\cite{gidea-tda-crypto, gidea2018topological,climate-ee-tda}. This is partly because these features are easier to visualize, interpret, and incorporate into standard data-analysis workflows, while higher-dimensional features have been less systematically explored. It is also because high-dimensional TDA is computationally expensive. For a complex on $N$ vertices, the number of possible $d$-simplices grows as ${N \choose d+1}$, leading to a combinatorial bottleneck as the homological degree $d$ becomes large.

Quantum algorithms offer a potential route around this bottleneck. The original algorithm for quantum TDA proposed by Lloyd et al.~\cite{lloyd2016quantum} described how to implement the large sparse matrices from algebraic topology as efficient quantum operators and measure the corresponding Betti numbers. This approach appeared attractive as its implementation requires only one qubit per graph node, regardless of the complexity of the data associated with that node, and it is able to avoid many of the pitfalls of data loading and readout which are often associated with algorithms in quantum machine learning. This work suggested the possibility of exponential speedups over classical methods under the stated access assumptions. Recent complexity theoretic results has made this picture more precise. While quantum advantage for Betti number estimation is not expected to be universal, there are regimes and graph families in which large quantum speedups are expected to persist \cite{berry2024analyzing,schmidhuber2023complexity}. There has also been recent progress in adapting key components of the algorithm for near-term quantum devices~\cite{akhalwaya2023topological}. The central question is therefore whether the favorable regimes overlap with data-analysis tasks of practical value.

In this work, we connect the dots on many of these ideas, and propose that a quantum algorithm for topological data analysis is both within reach of modern day quantum computers, and can provide value for important problems in data science.  We argue that quantum TDA should be viewed not only as a method for computing abstract topological invariants, but as a feature-extraction method for complex data analysis. In many applications, the goal is not to compute exact Betti numbers in the most general setting, but to extract topological information that improves prediction, classification, or detection of changes in an underlying system.  Our work represents an example of a bottom-up approach to quantum application development \cite{babbush2025grand}, where we first identify an algorithmic regime where quantum methods may provide an advantage, and then find applications where the information extracted may provide utility for real data analysis. 

To this end, we take a two pronged approach to evaluate the potential of a quantum algorithm to provide real world value through the TDA algorithm. First, we demonstrate the usefulness of high-order TDA features in two time series applications of practical relevance: functional MRI analysis for neuroedegenerative disease classification and financial time series analysis for detecting early indicators of market instability. One surprising idea is the fact that low-dimensional time series data taken from complex systems can contain information about the hidden dynamics, and may itself be represented as a higher dimensional network whose topology reflects aspects of those dynamics. We show this context that high-order TDA features are effective at revealing new information which can improve prediction in classical workflows.

Second, we develop a quantum approach for extracting the relevant topological features using low-order moments of the combinatorial Laplacian. This reframes the near-term quantum task from exact Betti number estimation to the extraction of Laplacian-derived observables that are strongly correlated with high-dimensional Betti information in the regimes studied here. Notably, we find that this approach does not appear to require the relative Betti number to be large, substantially expanding the class of simplicial complexes where quantum TDA may be useful. We provide resource estimates and quantum-classical crossover projections for this moment-based method, and demonstrate the algorithm on barium-based trapped-ion hardware by extracting topological observables from graph instances with up to 16 nodes.

The rest of this paper is organized as follows. In Sec.~\ref{sec:background}, we review the  mathematical background necessary for understanding TDA, including simplical complexes, Betti numbers, persistent homology and classical algorithms for computing topological features. We also analyze the resources required by state-of-the-art classical software, establishing a baseline for comparison with quantum TDA. In Sec.~\ref{sec:applications}, we dive into two applications where TDA can be used to study time series data: brain disease classification from functional MRI (fMRI) data, and calculating early indicators of market instability from financial time series.  In Sec.~\ref{sec:qTDA_theory}, we develop the quantum algorithm, analyze its
resource requirements, study the relationship between low-order Laplacian
moments and Betti numbers, and present quantum hardware experiments.  We find that contrary to conventional wisdom, the resources to perform this calculation do not depend strongly on the graph having a particularly large relative Betti number. Together, these results outline a clear pathway to potential quantum utility using the quantum TDA algorithm.

\clearpage 

\section{Background and Theory}\label{sec:background}

In this section, we give a brief overview of the important terms and mathematical background required to understand the main ideas in topological data analysis. We will define the idea of a simplicial complex as the generalization of a simple graph, and describe the ways these complex can be constructed from data. We also discuss how we can define Betti numbers as topological invariants of these complexes, and discuss the concept of persistence and persistent homology. In Sec.~\ref{sec:classical_alg}, we discuss the computational complexity of classical algorithms for TDA, and report numerical experiments on the computational resources required to perform these calculations using software packages developed for this purpose. This will set a baseline against which we can compare the performance of our quantum TDA algorithm. These simulations show how classical TDA methods struggle to compute high order Betti numbers, even on relatively small graphs and using significant classical resources.

\subsection{Terminology and Mathematical Foundations}

\begin{figure}[h]
    \centering
    \includegraphics[width=0.43\linewidth]{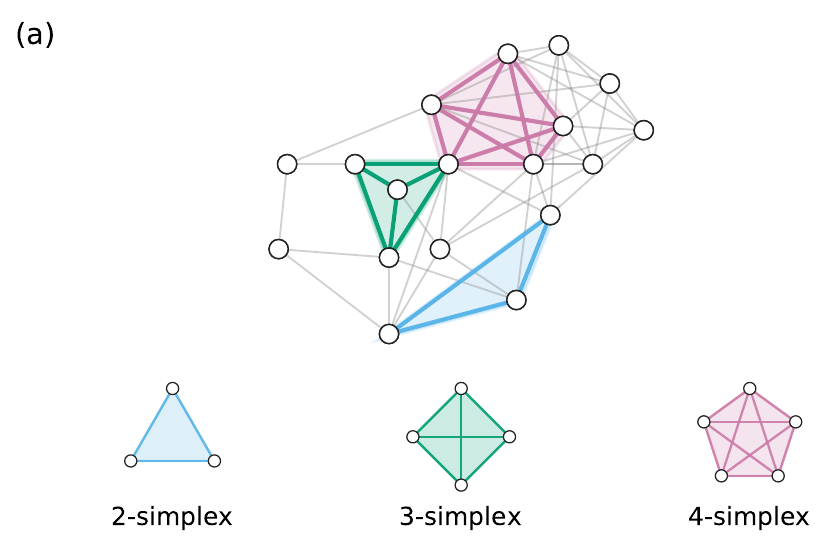}
    \quad \,\,
    \vline\,\,
    \quad
    \includegraphics[width=0.43\linewidth]{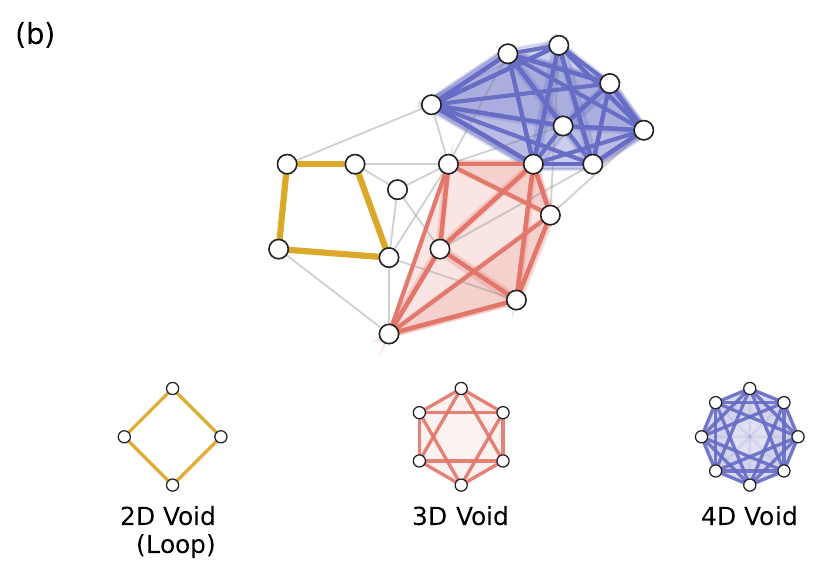}
    \caption{A simplicial complex on 18 nodes. We highlight subgraphs which form the basic $d$-simplices {\it left)}, and the simplex examples of 2D, 3D and 4D voids {\it (right)}.}
    \label{fig:k-simplex}
\end{figure}

 The main idea of this subject area is to provide a set of mathematical tools to study higher order networks, where interactions beyond simple pairwise or dyadic relationships are explicitly included in the model. To this end, the main object that we will consider is known as a {\bf simplicial complex}.  A simplicial complex may be thought of as a generalization of a graph, where we explicitly include geometric structures beyond nodes and edges. In particular, a simplicial complex is a set of nodes, edges, and generally d-simplices, for $d\in \mathbb{Z}$.  A {\bf d-simplex}, as shown in Fig.~\ref{fig:k-simplex} is a d-dimensional polytope that is the convex hull of $(d+1)$ vertices. In other words, a d-simplex is the generalization a triangle or tetrahedra to arbitrary dimension $d$. In a graph, the vertices and edges of a d-simplex form a {\bf k-clique} with $k=d+1$, that is a fully connected sub-graph on those k vertices.  So a simplicial complex, also shown in Fig.~\ref{fig:k-simplex}, includes the usual nodes and edges (0-simplexes and 1-simplexes respectively) which are typically associated with a graph, but also includes higher dimensional objects such as triangles (2-simplexes), tetrahedra (3-simplexes) and generally d-simplices. 

In a general network with $N$ vertices, the number of {\it potential} k-cliques for any order $k=d+1$ is given by $N \choose k$, and the number of total potential simplexes over all orders $k$ is $\sum_k {N\choose k+1} = 2^N-1$. Therefore, even enumerating all the elements of a simplicial complex can be computationally challenging. There are several different methods which can construct a simplicial complex from a given  set of points (nodes) in a space with an associated measure of distance between the nodes (i.e. a metric space). By far the most common structure, which we will use, is known as the {\bf Vietoris Rips (VR) complex}. In this construction, a k-clique is added to the complex if {\it all pairwise distances} between the set of k points are less than a specified threshold $\epsilon$. Therefore, an edge is added to the complex if the two endpoints are less than $\epsilon$ distance apart, a triangle is added if all edges of the triangle are $< \epsilon$ apart, a tetrahedron is added if all of its triangular faces are also included in the complex, and so on for higher order k-cliques. At a fixed value of $\epsilon$, a VR complex is also a {\bf clique complex}, where given a simple graph $G$, we form a complex by also including every d-simplex that is a {\bf clique} of the graph (i.e. a set of k nodes in G that are fully connected to each other). VR complexes have the advantage that they are defined using only pairwise distances. However, the number of simplices in the resulting clique complex can still scale as $\mathcal O(2^N)$ in the worst case. There also exist other complexes which can be useful for analyzing the underlying data, and have their own advantages and disadvantages \cite{otter2017roadmap}. One example is the \v{C}ech complex, where a d-simplex is included in the complex if and only if the intersection of all balls of radius $\epsilon/2$ centered around each of the $k=d+1$ points is non-empty.

\begin{figure}[H]
    \includegraphics[width=1.0\linewidth]{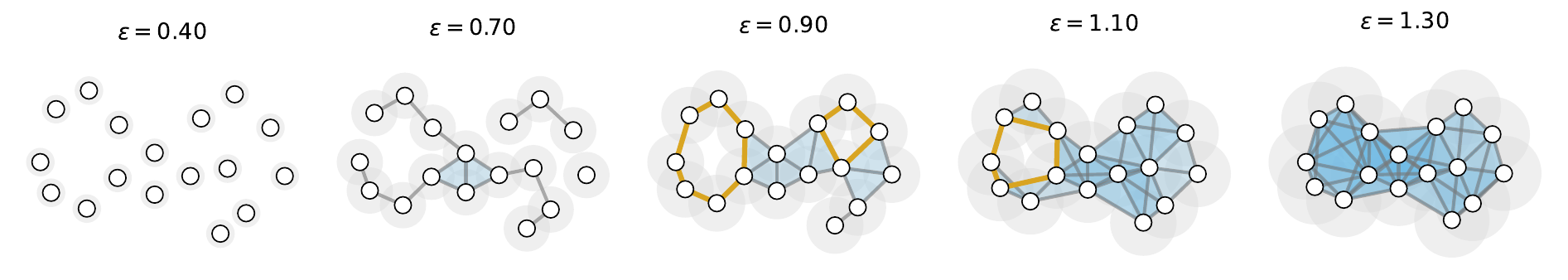}
    \caption{ A filtration showing the VR complex of a set of nodes at different length-scales $\epsilon$. At each length scale, edges are added between two nodes if they are separated by distance less than $\epsilon$. In the VR complex shown here, d-simplices are added if all $k=d+1$ nodes are connected with each other. }
    \label{fig:vr_filtration}
\end{figure}

The tool we use to analyze these simplicial complexes is known as {\bf homology}. Coming from the Greek word meaning ``same relation", homology assigns algebraic invariants to a space that summarize aspects of its shape. In TDA, these invariants can then be used to measure the degree of similarity between different data sets. In particular, we measure the topological properties of the simplicial complex formed by the data in the metric space. The topology of the complex measures the global characteristics of its shape, which remain the same under smooth local changes to the structure. Specifically, we measure the so called {\bf Betti number}, $\beta_d$, in dimension $d$, which informally speaking, counts the number of ``(d+1)-dimensional holes" in the complex. 

In fact, the main tool which is used to generate features that characterize a data set measures the homology as a function of length scale. This process gives a measure that combines both the local geometry and the global topology of the data set. To do this, a sequence of simplicial complexes are formed. A {\bf filtration} is a sequence of simplicial complexes, $\Gamma_i$, which can be nested within each other, so that $\Gamma_0 \subset \Gamma_1 \subset \dots \subset \Gamma_M$. When working with VR complexes, the filtration can be defined as a function of the distance parameter, $\epsilon$, as shown in Fig.~\ref{fig:vr_filtration}. This method is known as {\bf persistent homology}, as it tracks which topological features of the data set \emph{persist} across different length scales.



\subsection{Classical Topological Data Analysis} \label{sec:classical_alg}

\subsubsection{Classical Algorithm for Calculating Betti Numbers}

The $d^{\text{th}}$ Betti number of a simplicial complex measures the number of independent $d$-dimensional homology classes. Equivalently, it counts independent $d$-cycles that are not the boundary of a filled-in $(d+1)$-dimensional object. Intuitively, nontrivial d-cycles often correspond to voids one dimension higher: $\beta_0$ detects disconnected components, $\beta_1$ counts loops and $\beta_2$ counts enclosed voids or cavities. Mathematically, this can be defined using the tools of algebraic topology. We specifically define a series of operators which act on the space of simplicial complexes. The state space that we are working in is the set of all possible d-simplices which can be formed on $N$ nodes.  Each $d$-simplex (which is also a $k$-clique with $k=d+1$) may be represented by a bitstring of length N, with $k$ ones on the nodes which represent the vertices of the simplex. For example, a 2-simplex on nodes $\{0,2,4\}$ of a 7 node graph would be represented with the state $\ket{1010100}$. In this way, each k-clique subspace, $\mathcal{H}_k$ can be represented by states with Hamming weight $k$, and is of size $\binom{N}{k}$.  The full Hilbert space is then the direct sum $\mathcal{H}=\oplus \mathcal{H}_k$, and is of size $2^N$. Each simplex has an orientation assigned to it, so that the ordering of the nodes of the graph matters, with two simplices having the same orientation if they differ by an even permutation of the vertices. Note that throughout this work, we use $d$ for the homological dimension used in the standard TDA notation, giving $\beta_d$. We use $k$ for the Hamming weight of a computational basis state, equivalently the size of the clique represented by that state. Thus a state in $\mathcal H_k$ represents a $k$-clique, or a $(k-1)$-simplex, and corresponds to homological dimension $d=k-1$. Accordingly, the Laplacian acting on $\mathcal H_k$ has kernel dimension $\beta_{k-1}$.

We can now analyze the structure of the simplicial complexes by defining operators in this space. A particularly important example is the {\bf boundary operator} $\partial_k : \mathcal{H}_{k}\rightarrow \mathcal{H}_{k-1}$ and its adjoint $\partial_k^\dagger$. These operators connect k-cliques, or equivalently $(k-1)$-simplices, to their $(k-1)$-clique faces. In our Hilbert space we can define
\begin{eqnarray}
    \partial_k \ket{x} = \sum_{i=0}^{k-1} (-1)^i \ket{x\setminus i},
\end{eqnarray}
where $\ket{x\setminus i}$ is the basis state $\ket{x}$ with the $i^{\text{th}}$ $\ket{1}$ set to $\ket{0}$. This represents the operation of removing the $i^{\text{th}}$ vertex from the $k$-clique $\ket{x}$, and the sign keeps track of the orientation of the resulting $(k-1)$-cliques. 

There are two interesting properties of the boundary operator which we can highlight. The first is that this operator can be exactly represented in the Hilbert space $\mathcal{H}$ using fermionic creation and annihilation operators $a$ and $a^\dagger$ as
\begin{eqnarray}
    \partial &=& \sum_i a_i, \quad \partial^\dagger =\sum_i a_i^\dagger, \quad B = \partial + \partial^\dagger = \sum_i (a_i + a_i^\dagger) \\
    B^2 &=& \partial\partial^\dagger + \partial^\dagger \partial,
\end{eqnarray}
where the relative orientation of the simplices acts in exactly the same way as the fermionic minus sign introduced by the anti-commutation properties of the $a,a^\dagger$ operators. Note that $\partial_k$ is the component of $\partial$ that maps $\mathcal{H}_k$ to $\mathcal{H}_{k-1}$. We can also describe this operator on a specific complex $\Gamma$, whereby $\partial_k^\Gamma$ gives zero if the simplex $\ket{x}$ is not in the complex $\Gamma$. This can be implemented using the projection operator $P_\Gamma = \sum_{x\in\Gamma}|x\rangle\langle x|$, so that $\partial_k^\Gamma = P_\Gamma \partial_k P_\Gamma$.  This formulation of the boundary operator also implies the second important property, which is that for all states $\ket{x} \in \mathcal{H}_{k+1}$, we have
\begin{eqnarray}
    \partial^\Gamma_{k} \partial^\Gamma_{k+1} \ket{x} = \left[ P_\Gamma \left( \sum_i a_i \right) P_\Gamma \right ]^2 \ket{x} = 0.
\end{eqnarray}
This tells us that the image of $\partial_{k+1}^\Gamma$ lies in the kernel of $\partial_{k}^\Gamma$, and can be easily seen using the anti-commutation properties of the fermion operators. This property allows us to define the quotient space 
\begin{eqnarray}
    H_{d}(\Gamma) = \text{ker}(\partial_{d+1}^\Gamma) / \text{im}(\partial_{d+2}^\Gamma).
\end{eqnarray}
Mathematically, the Hamming-weight subspaces $\mathcal{H}_k$ subspaces along with the boundary maps $\partial_{k}$ define what is known as a chain complex, and the quotient space forms the homology group in degree $d=k-1$. These homology groups provide an algebraic characterization of the topology of the simplicial complex. Intuitively, $\ker(\partial_k^\Gamma)$ is the space of closed $d$-dimensional
structures: connected components for $d=0$, loops for $d=1$, closed surfaces for $d=2$, and higher-dimensional analogues. Meanwhile, $\text{im}(\partial_{k+1}^\Gamma)$ is the subspace of these closed structures that are boundaries of filled-in $(d+1)$-dimensional objects in $\Gamma$. The quotient therefore removes the cycles that can be filled in and keeps only the nontrivial topological features. The Betti number is then the dimension of this group, or the number of holes in the complex
\begin{eqnarray}
    \beta_{k-1} = \text{dim}(H_{k-1}(\Gamma)) = \text{dim}(\text{ker}(\partial_{k}^\Gamma)) - \text{dim}( \text{im}(\partial_{k+1}^\Gamma)).
\end{eqnarray}
The main goal of topological data analysis is to calculate the Betti numbers, thus characterizing the topology of the simplicial complex. 

One method of calculating these Betti numbers works by defining the Dirac operator
\begin{eqnarray}
    B = \begin{bmatrix}
        0 & \partial_{k}^{\phantom \dagger} & 0 \\
        \partial^\dagger_{k} & 0 & \partial_{k+1}^{\phantom \dagger}\\
        0 &\partial_{k+1}^\dagger &0
\end{bmatrix}.
\end{eqnarray}
Squaring this operator gives
\begin{eqnarray}
    B^2 = \begin{bmatrix}
        \partial_{k} ^{\phantom\dagger}\partial_{k}^\dagger & 0  & 0\\
        0 & \partial^\dagger_{k}\partial_{k}^{\phantom\dagger} + \partial_{k+1}^{\phantom\dagger}\partial_{k+1}^\dagger & 0 \\
        0 & 0 &\partial_{k+1}^\dagger \partial_{k+1}^{\phantom\dagger}
\end{bmatrix}.
\end{eqnarray}
The middle block of this matrix is an operator that lives in the subspace $\mathcal{H}_k$ and is known as the combinatorial Laplacian
\begin{eqnarray}
    \Delta_k^\Gamma  = \partial^{\Gamma\dagger}_{k}\partial_{k}^{\Gamma\phantom \dagger} + \partial_{k+1}^{\Gamma\phantom\dagger} \partial_{k+1}^{\Gamma\dagger},
\end{eqnarray}
and is the main operator that we will use in the quantum algorithm for TDA. We can construct $\Delta_k^\Gamma$ by constructing the $B^2$ operator restricted to the $\mathcal{H}_k$ subspace.  It is known that
\begin{eqnarray}
    \text{dim}(\text{ker}(\Delta_k^\Gamma)) = \beta_{k-1}.
\end{eqnarray}
We can therefore calculate $\beta_{d}=\beta_{k-1}$ by counting the number of zero eigenvalues of $\Delta_k^\Gamma$.

\subsubsection{Scaling of the Time-to-Solution of Classical Methods}
Here we analyze the scaling time to solution (TTS) of the classical TDA algorithm with the number of k-cliques, on Erd\H{o}s-R\'enyi random graphs. Each graph is composed of N nodes, with edges chosen at random to connect any two nodes. The graphs are then expanded to include higher dimensional cliques, forming a simplicial complex. In order to calculate the $d^{\text{th}}$ Betti number $\beta_d$ with $d=k-1$, the graph must be expanded to include cliques of order $k+1$. There are $\mathcal{O}\left(\binom{N}{k+1}\right)$ cliques of this size in such a complex. However, modern TDA software packages are able to use specialized techniques, such that the main bottleneck in the scaling scales only as $\binom{N}{d} = \binom{N}{k-1}$. In our simulations, we measure how the computational resources required to calculate these higher order Betti number scales. We verify that the TTS scales linearly with the number of (k-1)-cliques, as $\binom{N}{k-1}/(k-1)$, and the memory resources scale as $\binom{N}{k-1}$. Even with these advanced software packages, we see that the TTS becomes very large as the size of the graph grows.  Already for graphs of $\sim 80$ nodes, it becomes difficult to calculate Betti numbers above $d=8$.

\paragraph{With high performance multi-core multi-thread parallelization:}
We use the open source Python software package giotto-ph \cite{giotto-ph} which is particularly efficient for calculating persistent homology of Vietoris-Rips filtration.  It is built on existing tools like Ripser \cite{Ripser_Bauer_2021} and GUDHI \cite{gudhi:urm}, utilizing their efficient parallel thread implementation and the pre-processing edge-collapser algorithm to reduce the size of the simplicial complex. 

\begin{figure}[H]
\centering
{\includegraphics[width=0.49\linewidth]{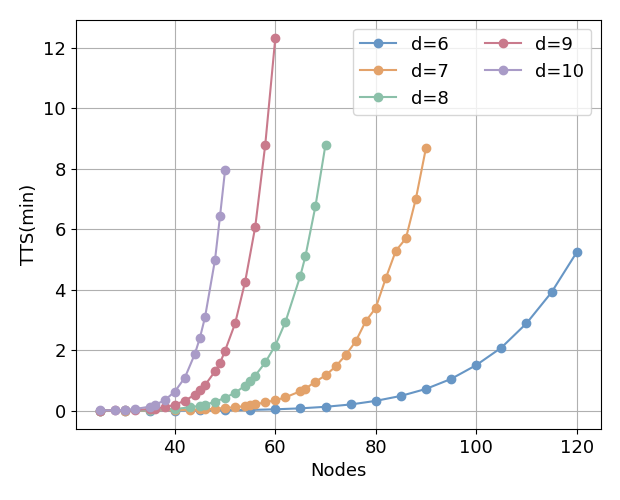}}~
\,\, {\includegraphics[width=0.40\linewidth]{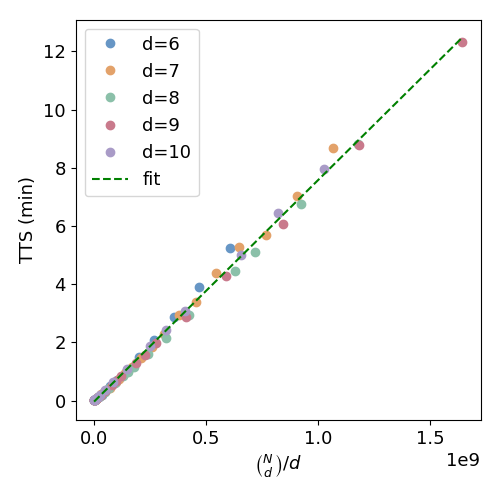}}
\caption{{\it (left)} TTS scaling with graph size for different $\beta_d$. {\it (right)} TTS scaling over all graphs and $\beta_d$: $\text{TTS} = a {N \choose d} / d + b$, where $a=7.6 \times 10 ^ {-9}, \ b=-0.037$.}
\label{fig:classical_TDA_resource_estimator_gph_1}
\end{figure}

We look at fully connected random graphs to estimate the TTS to calculate Betti numbers using giotto-ph on parallel CPU threads. In Fig.~\ref{fig:classical_TDA_resource_estimator_gph_1}, we show the TTS to calculate the k Betti number vs. number of graph nodes. We used up to 150 CPU nodes with 500 GB each and calculated up to $\beta_d$ with $d=10$. This represents our most efficient estimate on CPUs for calculating high-dimensional Betti numbers on random graphs. We also show a collapse of the TTS curves to extract the universal scaling function. We demonstrate that TTS scales linearly with $ {N \choose d} / d $ when using the giotto-ph package. Specifically, we find for $d=k-1$
\begin{equation} \label{eq: tts_scaling_giotto_ph}
    \text{TTS} = a {N \choose d} / d + b
\end{equation}
where, $a=7.6 \times 10 ^ {-9}, \ b=-0.037$. Note that this demonstrates the power of modern software packages to implement efficient algorithm for calculating Betti numbers, improving on the ${N \choose {k+1}}$ or even ${N \choose {k}}^3$ scaling that one might expect from a naive analysis of a TDA algorithm which measures the null-space of the combinatorial Laplacian. Nevertheless, the TTS values still explode even for relatively small graphs at high Betti numbers  

\begin{figure}[H]
\centering
{\includegraphics[width=0.40\linewidth]{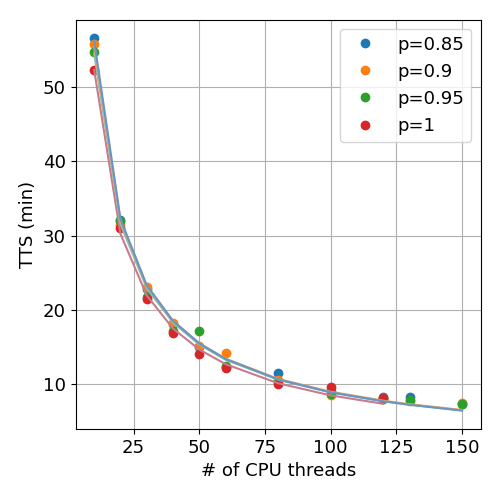}}~
\quad {\includegraphics[width=0.40\linewidth]{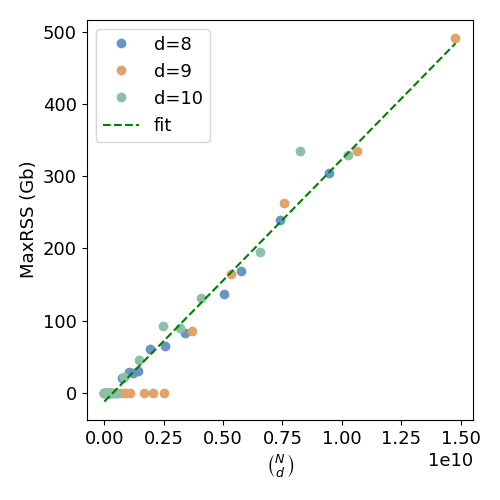}}
\caption{{\it (left)} TTS scaling with number of parallel CPU threads. {\it (right)} Memory scaling over all graphs and $\beta_d$: $\text{MaxRSS} = a {N \choose d} + b$, where $a=3.35 \times 10 ^ {-8}, \ b=-12.284$.}
\label{fig:classical_TDA_resource_estimator_gph_2}
\end{figure}

In Fig.~\ref{fig:classical_TDA_resource_estimator_gph_2}, we show TTS vs. number of parallel CPU threads. Here we calculate Betti= 10 for a random graph
with 50 nodes and different edge densities $p$. It can be seen that the computational resources can be effectively distributed among many CPU nodes. Notably, the TTS within giotto-ph does not depend strongly on the edge density of the graph. On the right plot, we show the universal scaling of the memory required to perform the Betti number calculation. In this case we find that the memory scales linearly with ${N \choose d}$. This is slightly worse scaling than the TTS shown in Fig.~\ref{fig:classical_TDA_resource_estimator_gph_1}, and therefore the memory requirements may present an earlier barrier to the classical algorithm than the total time.


\newpage
\section{On Applications to Commercially Relevant Problems}
\label{sec:applications}

To assess the value of quantum TDA, it is important first to examine the impact of higher-dimensional homologies, as their calculation represents the core technological advantage.
We build our work upon prior studies that applied classical TDA to financial time series and fMRI-based disease detection, which were previously restricted to $H_0$ and $H_1$ homologies. We expand these analyses by computing homologies up to $H_4$ on these established datasets to evaluate their utility for prediction tasks.


\subsection{Quantum TDA for Brain Function Analysis using FMRI Data} \label{sec:brainfmri}

Understanding complicated and changing connections in the human brain has been the focus of the National Institute of Health since 2011, when it launched the Human Connectome Project to characterize brain structure and study underlying neurological problems~\cite{connectome_url, essen-hcp-2012, hcp-nature-2016, hcp-review-2021}. Recent studies have shown that functional magnetic resonance imaging data can be used to form highly connected graphs that show functional relationships between many spatially separated nodes in the brain, leading to complexes where the existence of higher-dimensional structures may be more prevalent~\cite{lowe-fmri-2016, fmri-neuron-2017, rogers-fmri-2017,songdechakraiwut2021}. Recently, there is a growing body of evidence showing that taking the higher-order structure into account when modeling brain function can greatly enhance our modeling capacities and help us to understand and predict their emerging dynamical behaviors~\cite{bick2023higher,bullmore-nature-2009,Gracia-Tabuenca751008,kocevar-graph-theory-2016,rubinov-sporns-2010,sporns-2018,battiston2020,braintdatutorial}. Motivated by these studies and our quest for searching utility of higher-order homologies and Betti numbers, we focus on exploring fMRI data as a test bed for topological data analysis in this section.

We use \cite{bhattacharya_2024} and \cite{alarjani_almarri_2025} as references for existing state-of-the-art results in predicting brain disease from fMRI data. In Ref.~\cite{bhattacharya_2024}, the authors use persistent homology from Vietoris-Rips filtration to detect different stages of cognitive impairment in patients from their brain fMRI time series data. In particular, they construct a distance matrix from persistent diagrams of each pair of regions of interest (ROIs) for each patient, which is then used with a classical convolutional neural network (CNN) pipeline to predict the stage of cognitive impairment. They used fMRI time series from the publicly available ADNI dataset~\cite{adni2010} and Dosenbach atlas to get region specific ROIs. There are $160$ Dosenbach ROIs, which are categorized into $6$ distinct brain regions. They perform the full classification pipeline for each of the $6$ regions. Their best model on the ADNI dataset shows $\sim 81 \%$ accuracy (see Table.~\ref{tab:comparative_results}). Note that the ADNI dataset has $162$ EMCI (early mild cognitive impairment), $141$ LMCI (late mild cognitive impairment) and $177$ HC (healthy cognition) patients.


\begin{table}[h!]
  \centering
  \label{tab:comparative_results}
  \begin{tabular}{l@{\qquad}l@{\qquad}l@{\qquad}c@{\qquad}r@{\qquad}c}
    \hline 
    \hline
    \textbf{Source} & \textbf{Data} & \textbf{Dataset size} & \textbf{Brain region} & \textbf{Features} & \textbf{Accuracy} \\
    \hline
    
    \multicolumn{6}{c}{\rule{0pt}{16pt}\textbf{based on conventional feature extraction}} \\
    Ref.~\cite{alarjani_almarri_2025} & OASIS & $101$ AD, $102$ healthy &  Full &  & $86.16\%$ \rule{0pt}{16pt}\\
     & ADNI & $55$ AD, $95$ healthy$^{*}$ & Full & & $93.9\%$ \rule{0pt}{16pt}\\[2ex]

    \multicolumn{6}{c}{\rule{0pt}{16pt}\textbf{based on TDA-only classification}} \\[1ex]
    Ref.~\cite{bhattacharya_2024} & ADNI & $141$ AD, $177$ healthy & DMN & $H_0$ & $74.6\%$  \\
     &  &  & FP & $H_1$ & $75.4\%$\\
     &  &  & OP & $H_0$ & $81\%$ \\
     &  &  & SM & $H_0$ & $69.9\%$ \\
     &  &  & CO & $H_1$ & $77.3\%$ \\
     &  &  & CB & $H_2$ & $71.6\%$ \\[2ex]

    Our work & OASIS & $32$ AD, $79$ healthy & Full & $H_0$ & $69\%$ \\
     &  &  &  & $H_0:H_1$ & $70.5\%$ \\
     &  &  &  & $H_0:H_2$ & $72.5\%$ \\
     &  &  &  & $H_0:H_3$ & $73.5\%$ \\
     &  &  &  & $H_0:H_4$ & $74\%$ \\
    \hline
    \hline
  \end{tabular}
  \caption{Comparison between related benchmarks and methods to perform classification. The reported accuracies are not directly comparable as the datasets, pre-processing pipelines, atlasses, and model classes differ across studies. They are included to provide context for the TDA-only classification results in this work. Note also that the TDA-only methods can in principle be used to enhance conventional feature extraction. $^{*}$In Ref.~\cite{alarjani_almarri_2025} (ADNI), the dataset was further augmented by SMOTE methods.}
  \label{tab:comparative_results}
\end{table}

In contrast, Ref.~\cite{alarjani_almarri_2025} uses fMRI data from both the OASIS and the ADNI datasets to perform a similar classification task. However, they use correlation matrices between inter-ROI time series as features, instead of TDA based ones. Their OASIS dataset has $101$ Alzheimer's stage, $95$ MCI (mild cognitive impairment) stage and $102$ healthy individuals. They perform binary classification between each pair of the categories and a multi-class classification. In addition, they used advanced data pre-processing techniques distinct from our approach and used the AAL3 (Automated Anatomical labeling) atlas to separate into $170$ ROIs~\cite{aal3_atlas_ref}. They use non-iterative ensemble approaches like extreme learning machines (ELM) for classification, where multiple ELMs are trained on different subsets of the features and the predictions are aggregated. They report an average accuracy of $\sim 86\%$ on classifying between Alzheimer's diseased (AD) and healthy patients from the OASIS dataset.


In our study, we used only fMRI data from the OASIS dataset. 
We first pre-process the data (scan videos) to make it compatible with our TDA pipeline. Each scan file is a $4$ dimensional data, which represents a time series associated with each 3D voxel of the brain due to its varying signal intensity over time. The scan files are in NIFTI (Neuroimaging Informatics Technology Initiative) format which are converted to an array format with appropriate python packages~\cite{Nilearn}. 
The performance of the classification neural network is influenced by several factors. A crucial initial step is the identification of the Regions of Interest (ROI), or major cluster nodes in the brain ~\cite{liu_roi_2011, poldrack_roi_2007}. This step significantly compresses the total size of the connectivity graph from millions to a few hundred nodes while retaining rich information about functional connectivity.
Another significant factor is the dataset size. Our dataset, while utilizing the same OASIS dataset as Ref.~\cite{alarjani_almarri_2025}, is composed of fewer individuals.
Unlike the approach in Ref.~\cite{alarjani_almarri_2025}, we employ embedding techniques to convert time series data into high-dimensional point clouds. This transformation necessitates a sufficient number of time stamps in each time series so that every patient have similar sized point clouds. Note that the fMRI time series of all patients were not equal. Hence, to ensure consistency across all patients, we finally have $32$ AD and $79$ healthy individuals whose resulting point clouds were of similar sizes. A true baseline model to benchmark against would be a model trained on this dataset with $111$ patients using a non-TDA method for feature extraction. For now, we list the results from Refs.~\cite{bhattacharya_2024} and \cite{alarjani_almarri_2025} for comparison.

We emphasize that this comparison should not be interpreted as a state-of-the-art benchmark for fMRI-based Alzheimer's classification. The datasets, atlases, pre-processing pipelines, feature representations and model classes all differ substantially across these studies. A true apples-to-apples comparison would require training conventional correlation-based or deep-learning models on the same reduced OASIS subset under the same validation protocol used here. Instead, the goal of this study is to test whether higher-dimensional TDA features contain disease-relevant signal beyond the information captured by low-dimensional homology features alone. 

\begin{figure}[H]
    \centering
    \includegraphics[width=0.4\linewidth]{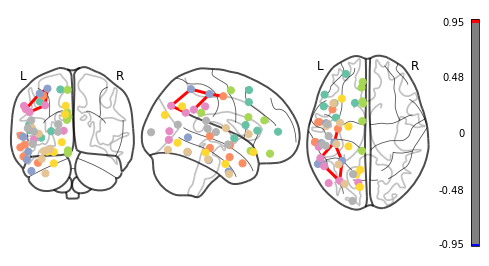}
    \includegraphics[width=0.4\linewidth]{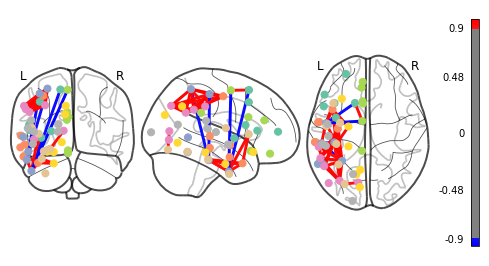}

    \includegraphics[width=0.4\linewidth]{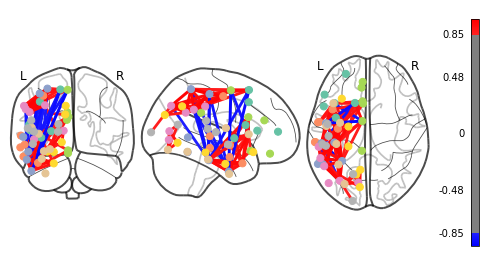}
    \includegraphics[width=0.4\linewidth]{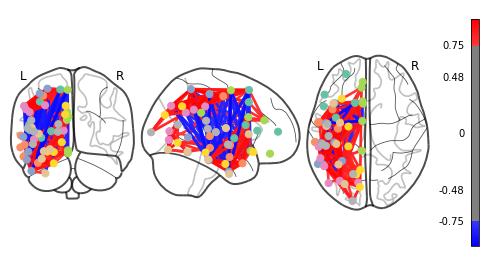}
    
    \caption{Connectome evolution over different thresholds on the correlation matrix from a patient's data from the OASIS dataset. {\it (top left)} Threshold $=0.95$, {\it (top right)} threshold $=0.9$, {\it (bottom left)} threshold $=0.85$, {\it (bottom right)} threshold $=0.75$.}
    \label{fig:connectome_sample}
\end{figure}

We performed the ROI extraction with a python tool called Nilearn~\cite{Nilearn} and used the Harvard-Oxford atlas to identify 48 cortical structural areas in each fMRI data. For each scan we used the built-in covariance estimator to extract a $48 \times 48$ dimensional correlation matrix. As shown in Fig.~\ref{fig:connectome_sample}, this correlation matrix allows for the construction of graphs with varying connectivity by applying different thresholds.



\subsubsection{Point Clouds from Regions of Interest (ROI) Time Series} \label{sec: pc_to_ts}



\begin{figure}[H]
    \centering
    \includegraphics[width=0.92\linewidth]{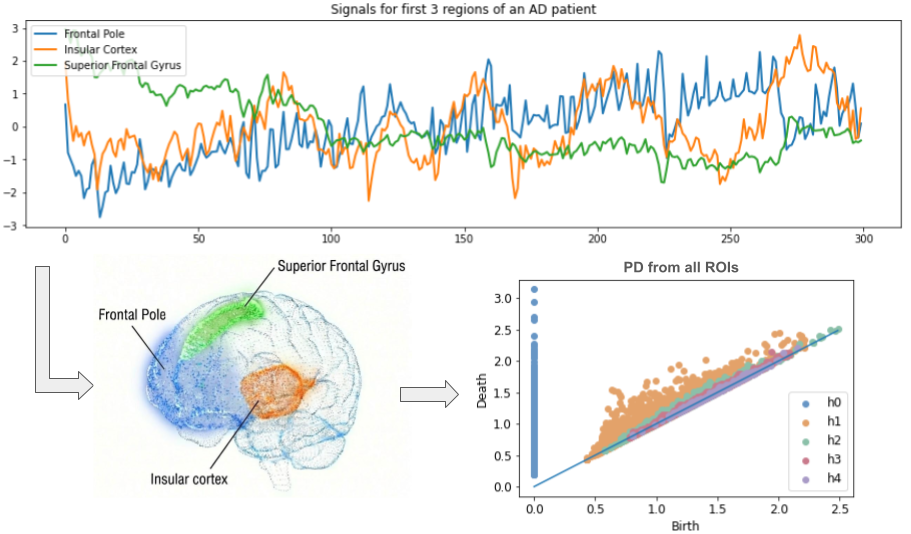} 
    \caption{ROI time series, point clouds and persistent homology on a sample patient data. {\it (top)} Time series from first 3 ROIs (using the Harvard-Oxford atlas) of an individual. {\it (bottom left)} Representative point clouds created by embedding techniques from each ROI time series. {\it (bottom right)} Persistence diagram from the point clouds.}
    \label{fig:ts-to-pd}
\end{figure}

Motivated by Ref.\cite{bhattacharya_2024}, we extract time series from each of the $48$ ROIs, and convert them to $10$ dimensional point clouds with $\sim 150$ points using Takens embedding from the giotto-tda library~\cite{giottotda}. The authors in Ref.\cite{bhattacharya_2024} provide evidence that this approach gives better performance than leveraging correlation matrices for graph construction. Next, using Vietoris-Rips filtration, we extract persistent homology information on these point clouds for homology dimensions $0$ to $4$. This gives birth and death information of all features in each dimension across different values of the filtration parameter, shown in the bottom panel of Fig.~\ref{fig:ts-to-pd}. 



After extracting the persistence diagrams (PDs) for each ROI, we compute the distance between every pair of ROI PDs, generating a distance matrix for each homology dimension (as detailed in Eq.~\ref{eq: dist_mat_PD}). This matrix encapsulates rich information about the functional connectivity between different regions of the brain. When this methodology was applied to fMRI data from two patients -- one cognitively healthy and one diagnosed with Alzheimer's -- it successfully distinguished between the functional connectivity patterns of the two brains.

\begin{equation} \label{eq: dist_mat_PD}
\begin{pmatrix}
D(PD(ROI_1, ROI_1)) & D(PD(ROI_1, ROI_2)) & \cdots & D(PD(ROI_1, ROI_n)) \\
D(PD(ROI_2, ROI_1)) & D(PD(ROI_2, ROI_2)) & \cdots & D(PD(ROI_2, ROI_n)) \\
\vdots & \vdots & \vdots  & \vdots\\
D(PD(ROI_n, ROI_1)) & D(PD(ROI_n, ROI_2)) & \cdots & D(PD(ROI_n, ROI_n))
\end{pmatrix}
\end{equation}
\vspace{0.3cm}

\begin{figure}[H]
    \centering
    \includegraphics[width=0.65\linewidth]
    {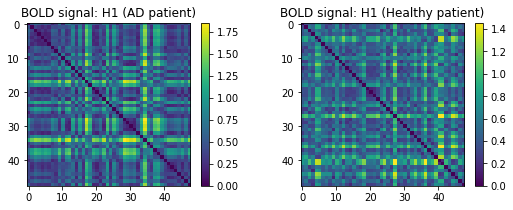}
    \\
    \includegraphics[width=0.65\linewidth]
    {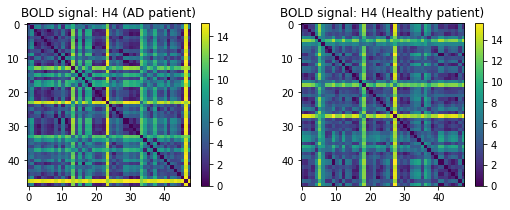}
    \caption{Features from blood oxygen level (BOLD) signals: Distance matrices between persistence diagrams of brain ROIs of a sick and a healthy patient. {\it (top panel)} $H_1$ features, {\it (bottom panel)} $H_4$ features.}
    \label{fig: dist-mat}
\end{figure}

In Fig.~\ref{fig: dist-mat}, we show distance matrices between the two patients using mb4-BOLD data fMRI data. MB4, or multiband factor 4, is a setting used in resting state fMRI to analyze different brain functions through blood oxygen level dependent (BOLD) signal. Through BOLD signals, researchers can observe changes in metabolic activity through neurological signals.

\subsubsection{Using fMRI-TDA Features for Disease Diagnosis}
The TDA features extracted above are next integrated with a classifier model to identify individuals who are in the AD stage. We trained two separated classifiers: a neural network (NN) and a Support Vector Machines (SVM) based model, and demonstrated that when higher homology features are included, the classification accuracy improves substantially compared to when only lower homological features such as $H_0$ or $H_0+H_1$ are used. 

The purpose of these classifier experiments is to isolate the contribution of higher-order topological features. In particular, we focus on whether performance improves as homology dimensions beyond $H_0$ and $H_1$ are added. This allows us to test whether the higher-dimensional TDA features themselves carry useful signal, rather than merely increasing the input dimension of the classifier.

\begin{figure}[H]
    \centering
    {\includegraphics[width=0.3\linewidth]{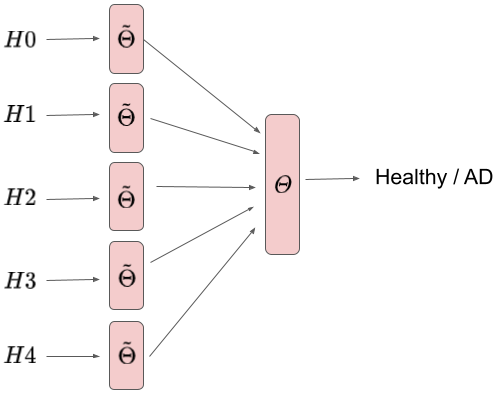}}~
    {\includegraphics[width=0.3\linewidth]{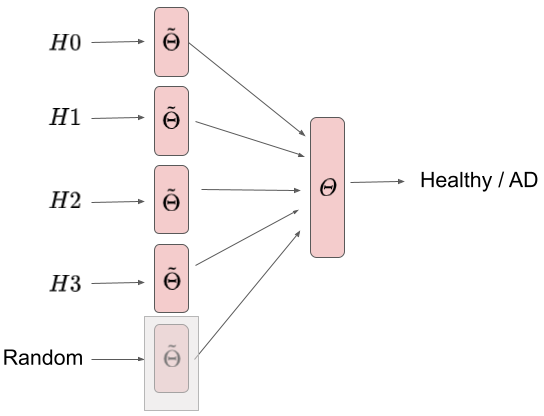}}~
    {\includegraphics[width=0.3\linewidth]{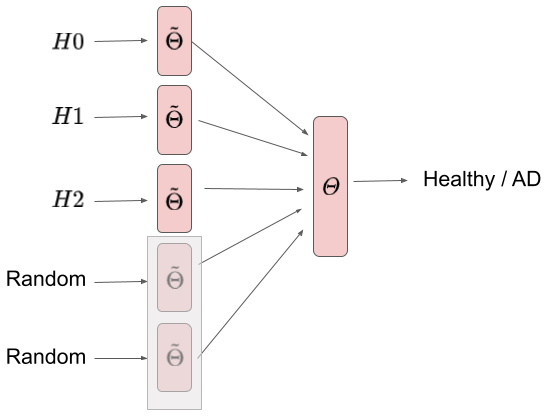}}
    \caption{Comparing different scenarios for feature extraction: {\it (left)} utilizing all five TDA features; {\it (middle)} employing four TDA features, with the last one substituted with a random feature; and {\it (right)} using three TDA features, with the remaining two replaced by a random feature. Note that the first layer of the Neural Network (NN) blocks shares parameters across these scenarios.} 
    \label{fig:aggregated_NN_model}
\end{figure}

\noindent \textbf{Neural network classification results:} 
A simple two-layer linear neural network (NN) was trained to predict the health status (sick or healthy) of patients using fMRI data. The dataset comprises 79 healthy and 32 sick patients.
The NN model utilizes homology features, ranging from $H_0$ to $H_4$, as inputs. The architecture, illustrated in Fig.~\ref{fig:aggregated_NN_model}, consists of a first layer with parameters $\tilde{\theta}$ that processes the homology features, and a final layer with parameters $\theta$ that aggregates these outputs to make the prediction. Notably, all homology features are fed into the same first layer ($\tilde{\theta}$).
To maintain a consistent number of trainable parameters across different homology dimensions, any unused features are replaced with random features. For instance, when only $H_0$, $H_1$, and $H_2$ features are used as inputs, the $H_3$ and $H_4$ features are substituted with random values (as depicted in the middle panel of Fig.~\ref{fig:aggregated_NN_model}).

The TDA features (see Fig.~\ref{fig: dist-mat}) are high dimensional $48\times48$ matrices. Using these features without dimensionality reduction will require large neural networks which are prone to overfitting. Consequently, we compressed the features to arrays of size $12\times 12$ in order to train smaller neural networks and reduce the effect of overfitting. The NN training was performed using a $80\%-20\%$ train-test split. Note that the dataset is inherently imbalanced, and we maintain the ratio of healthy to sick patients in the train-test split of the dataset.

\begin{figure}
    \centering
    \includegraphics[width=0.95\linewidth]{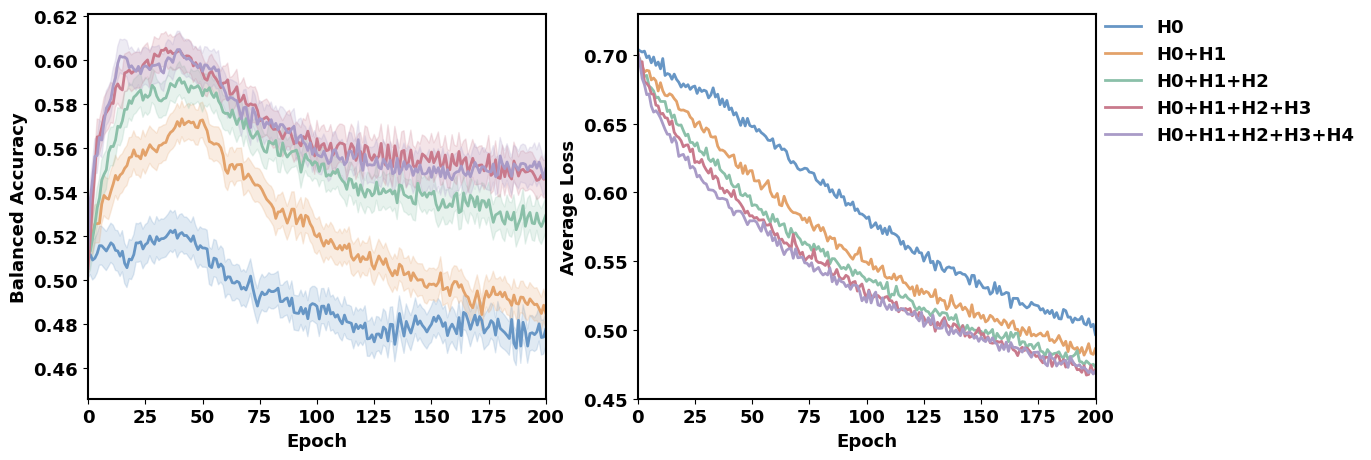}
    \caption{Balanced accuracy on test data {\it (left panel)} and training loss {\it (right panel)} for NN models trained using different combinations of homology features. The balanced accuracy scores and training losses are averaged over 200 independent runs.  }
    \label{fig:NN_Training_01}
\end{figure}

\begin{figure}
    \centering
    \includegraphics[width=0.8\linewidth]{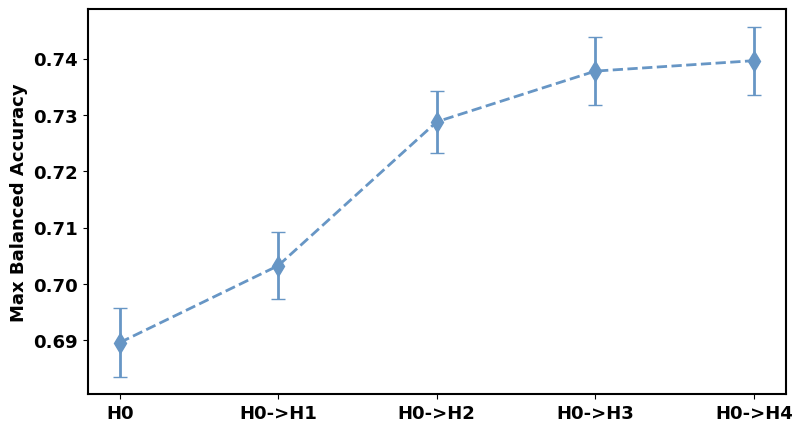}
    \caption{Average maximum balanced accuracy scores for NN models trained using different combinations of homology features. }
    \label{fig:NN_Training_1}
\end{figure}

We trained NN models using different homology features as inputs. Since the dataset is imbalanced, we use the balanced accuracy metric (on test data) to compare the performance on the models. Figure \ref{fig:NN_Training_01} shows the evolution of the average balanced accuracy (left panel) and average loss (right panel) during training. The averages are obtained from 200 independent runs. From the figure, we see that models trained using higher dimensional homology features perform better than those trained using lower dimensional homology features. This result indicates that higher dimensional homology features could be more effective in classification tasks that use fMRI data. 

We can also assess model performance by examining the average of the maximum balance accuracies reached during training. Note that for a model trained using a particular set of homology features, the maximum balanced accuracies of the 200 independent runs could be reached at a different epochs. Figure \ref{fig:NN_Training_1} shows the average of the maximum balanced accuracies for NN models trained using different homology features. Similarly to the previous case, this figure demonstrates that models trained using higher dimensional homology features outperform those trained using lower dimensional homology features. The model trained using all the homology features achieved the maximum balanced accuracy which is around 74\%. We believe that this is a satisfactory performance despite the small dataset size and significant class imbalance.

\textbf{Support Vector Machines (SVM) classification results:}  We also used the SVM algorithm to train classification models using the TDA features. The TDA data was pre-processed as in the previous case using compression sizes 2,4,6, 8, and 10. The dataset used for training and validation has 79 healthy patients and 32 sick patients. We used 80\% of the dataset for training and the rest for testing. As is it was in the previous case, we kept the ratio of healthy to sick patients in the test dataset the same as the ratio in the entire dataset. We trained SVM models using stacked homologies data as features. The models were trained using 100 random train-test splits of the dataset, and the results are presented in this section. 

\begin{figure}[h]
    \centering
    \includegraphics[width=0.9\linewidth]{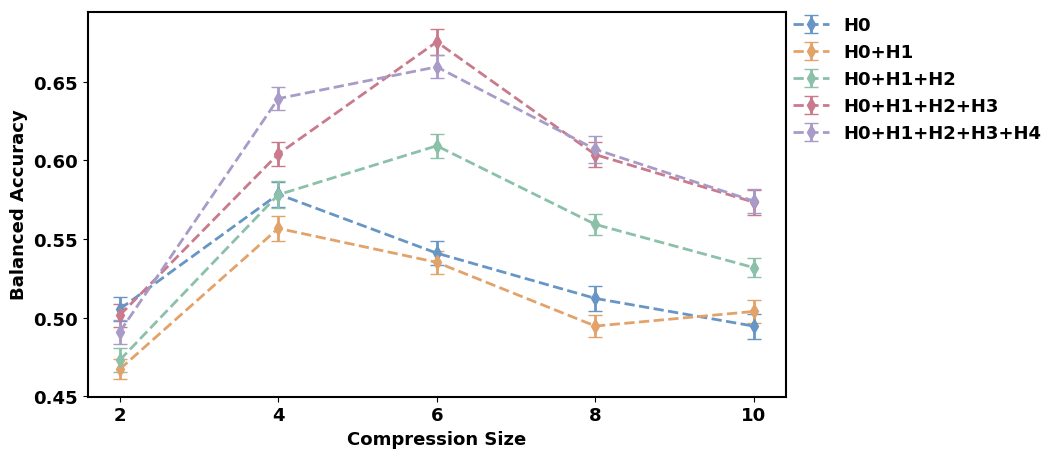}
    \caption{Average balanced accuracy on test dataset versus compression size for SVM models trained using stacked homology features. The dataset used for training and testing has 79 healthy individuals and 32 sick individuals.}
    \label{fig:SVM00}
\end{figure}

Figure \ref{fig:SVM00} shows the test average balanced accuracy vs compression size for each of the SVM models trained. From the figure, we do see that for all compression sizes, models trained using higher homology features have better performance than those trained using lower homology features. This result is another indication that features from higher homology dimensions might be more useful for Alzheimer's classification using fMRI data. The maximum SVM performance reached is about 66\% (compression size = 6) which is comparable to the best NN models (see Fig.~\ref{fig:NN_Training_01}). Also, the average balanced accuracy decreases as the compression size increases beyond 6, which is likely caused by over-fitting.

Overall, these results provide preliminary evidence that higher-dimensional TDA
features extracted from fMRI time series contain information related to disease
state. The absolute classification accuracy remains below the strongest
reported fMRI pipelines, and a complete benchmark would require comparison to
non-TDA models trained on the same reduced OASIS subset. Nevertheless, the
consistent improvement observed when adding higher homology dimensions suggests
that these features encode meaningful information that is not captured by
lower-dimensional TDA features alone. This is the central point for the present
work: the high-order topological information targeted by quantum TDA appears to
be useful in a real biomedical time series dataset.

paragraph{Summary.}
In this section, we demonstrate that topological features from fMRI scans provide useful insights about brain health. In particular, we showed that TDA features can be used in integration with a ML pipeline to perform classification between Alzheimers and cognitively healthy patients, and that higher homological TDA features improved performance on our dataset. This result is relevant for quantum TDA because these higher homologies can become exceptionally difficult to compute classically on large, densely connected graphs, such as those arising from more detailed brain atlas maps with hundreds of ROIs or large neuronal spike train datasets Ref.~\cite{ayhan2025persistenthomologypipelineanalysis, maslennikov2024spiketrain}. This is precisely where we expect the quantum TDA algorithm to perform more efficiently than currently available classical methods, thus paving the way for a better diagnostic framework for cognitive diseases.

Overall, these results provide preliminary evidence that higher-dimensional TDA
features extracted from fMRI time series contain information related to disease
state. The absolute classification accuracy remains below the strongest
reported fMRI pipelines, and a complete benchmark would require comparison to
non-TDA models trained on the same reduced OASIS subset. Nevertheless, the
consistent improvement observed when adding higher homology dimensions suggests
that these features encode meaningful information that is not captured by
lower-dimensional TDA features alone. The central takeaway is that that the high-order topological features targeted by quantum TDA appear to carry useful signal in a real biomedical time series dataset.

\subsection{Quantum TDA for Time Series Analysis of Financial Data} \label{sec:finance}

Financial institutions process large amounts of data in different ways and for a variety of purposes, such as forecasting market behavior and optimizing portfolios. The set of tools used for these calculations spans the range from heuristic indicators to stochastic simulations and complex machine learning models~\cite{Kaczmarek2022,portfolio-optim-AI,forecasting-methods-2018}. However, there are major challenges when working with the types of noisy time series data often associated with financial markets. For one, the market is a highly complex system with complicated high-order interactions between different assets affecting the overall behavior. Additionally, the data available for many tasks may be noisy, sparse, and difficult to obtain~\cite{fi-big-data,fi-data-complexity-2024}. These challenges can hinder attempts to understand the collective behavior of the market and create a need for more powerful tools for analysis. Predicting so called black swan events, such as extreme market swings, is a particularly difficult task of great importance\cite{ee-finance, ee-time-discretization-2026, ee-economics-price-bubbles, ee-risk-management}. In this section, we will demonstrate how techniques from topological data analysis, when applied to financial data, can effectively reveal hidden relationships between assets which evolve over given time periods.

Generally speaking, TDA tools can be used to diagnose structural changes in the market as a whole, which are expected to appear in the build up to a crash~\cite{gidea2018topological, fi-forecast-tda-2025, tda-invest-goel-2020, ee-stock-market-2024}. In Ref.~\cite{gidea2018topological}, it was seen that TDA can be effective at predicting extreme market events. Specifically, it was seen that the magnitude of persistent topological features hidden in the multi-dimensional time series data of the daily returns of four major US stock indexes produces a clear signal preceding a market crash. We further explore the connection between extreme events in the stock market and topological signals in financial time series data. We expand on this work to include higher order topological features, and find that these higher order features can produce stronger and more predictive signals of these extreme events. 

\subsubsection{Data Processing}
In order to analyze time series data using methods from TDA, the time series must be converted to a point cloud. The typical method for analyzing time series data as a point cloud is to use a {\it time delay embedding}, where points in a $D$-dimensional space are constructed from a single time series by sampling the time series at $D$ different times, each separated by a constant delay $\tau$. That is, we construct a series of points 
\begin{eqnarray}
{\bf y_t} = (y(t),y(t+\tau),y(t+2\tau),\dots,y(t+D\tau)).
\end{eqnarray}
A point cloud can then be constructed from the original data by generating multiple points at different locations, $t$, along the time series. {\it Takens embedding theorem} states that, as long as the dimension $D$ is sufficiently large, the resulting point cloud reconstructs the manifold of the attractor of the original dynamics. That is, the dynamics of the time series may generally depend on a large number of variables, $n$, and the time evolution of the complex system forms a geometric structure in the $n$ dimensional phase space. However, we can only measure specific observables, such as closing price as a function of time, which depend on these $n$ hidden variables. Takens theorem states that the shape of the full structure in the original n-dimensional phase space can be reconstructed just from time-delayed samples from the single variable, as long as we sample from enough points $D$. Therefore, the shape of the time-delay embedded point cloud contains important information about the relationship between hidden parameters of the dynamics. 

This method can be extended to cases where we have multi-dimensional time series data. In fact there are several different ways to construct a point cloud from multidimensional time series data. For example, the dimension of the points in the point cloud can be extended when we have $M$ correlated time series, to generate a point cloud in $MD$ dimensional space, by applying a time delay embedding which samples $M$ dimensional data at $D$ separate times. That is, 
\begin{eqnarray}
    {\bf y_t} = (y_0(t),y_1(t),\dots y_M(t)).
\end{eqnarray} 
This is the approach taken in Ref.\cite{gidea2018topological}, which has proven effective in measuring structural changes in the relationships between different indexes. On the other hand, this approach has limitations in extending it to higher order homologies as the number of points in the point cloud is limited by the length of the time series, and the dimension of the data is limited by the number of stocks being considered. This approach also does not take advantage of the results from Taken's embedding theorem mentioned above. 

Alternately, one may generate a separate point cloud for each of the M marginal time series functions, to get $M$ separate points clouds, each with $N$ points in $D$ dimensional space. One can then concatenate these point clouds together to form one larger point cloud with $NM$ points in $D$ dimensional space. That is, for each index $i$, we embed the time series as
\begin{eqnarray}
{\bf y_{i,t}} = (y_i(t), y_i(t+\tau),\dots y_i(t+D\tau)).
\end{eqnarray}
for stock index $i$ at time window starting at $t$.
In this work, we use this second approach to analyze the behavior of the stock market over a period of time between 2005 and 2024. We show our setup in Fig.~\ref{fig:Finance_tda}. We analyzed the closing stock price of a set of $17$ different indexes, over a 10 year period of returns, with each time window covering a period of 100 days of returns. We embed the data in 10 dimensions, with a delay $\tau$ of 2 days and a stride length of 1 day.  

\begin{figure}
    \centering
    \includegraphics[width=0.82\linewidth]{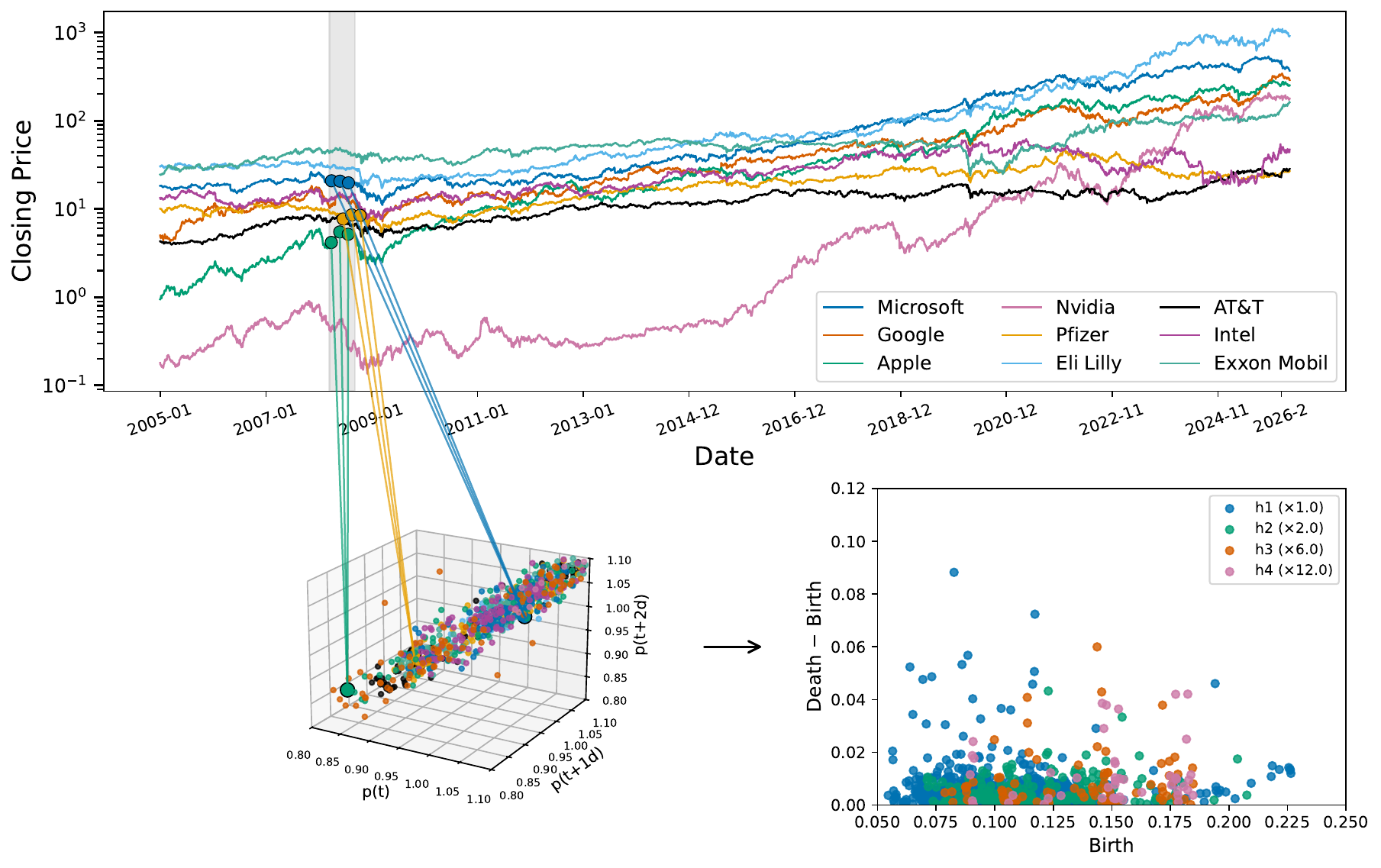}
    \caption{ Time-delay embedding of financial time series. Closing-price time series from multiple market indices are converted into a point cloud, where each point represents prices sampled from one index over a fixed time window. Persistent homology is then computed on this point cloud to produce a birth-death diagram summarizing its topological features.}
    \label{fig:time-delay-diagram}
\end{figure}

\subsubsection{Analysis of TDA Features}

\begin{figure}[H]
    \centering
    \includegraphics[width=0.94\linewidth]{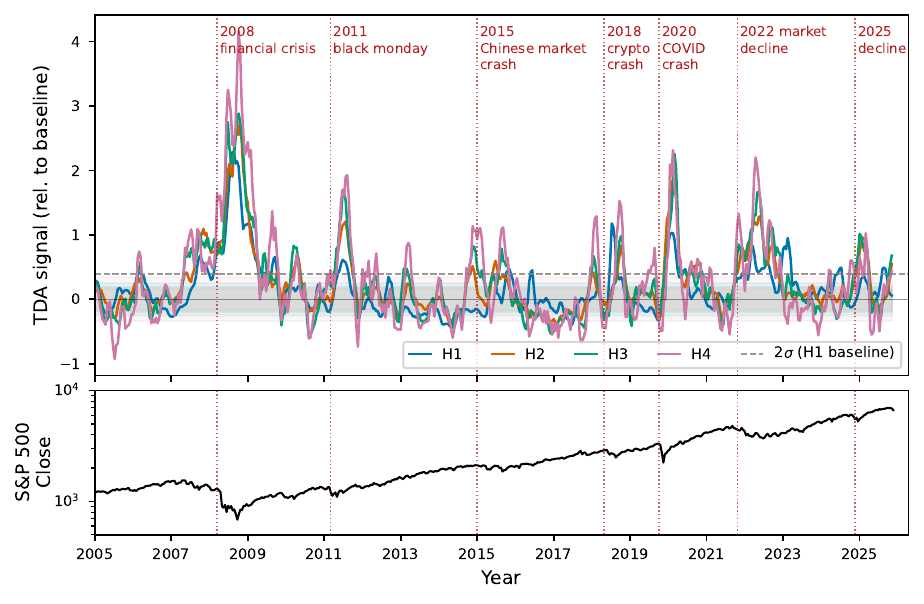}
    \caption{ The persistence curve for homology groups $H_1$ to $H_4$ of point clouds generated using a time-delay embedding of the closing stock price of 17 stocks in a time window of 100 surround the date on the x-axis. The corresponding closing value of the S\&P 500 index is also shown. Peaks in the TDA signals correlate with large downturns in the market value. Several large peaks are visible in the $H_4$ curve (green) which are smaller or absent in the low dimensional $H_1$ curve (blue).}
    \label{fig:Finance_tda}
\end{figure}

To measure the topological properties of the point clouds described above, we calculate the persistent homology of the point cloud of a given time window by generating a birth death diagram, like that shown in Fig.~\ref{fig:time-delay-diagram}. We then plot the average persistence for each time window as the $L_2$ norm of the birth-death diagram
\begin{eqnarray}
    L_2 = \sum_i |\lambda_i(d_i) - d_i|^2
\end{eqnarray}
where $\lambda_i$ is the death value of the i$^{\text{th}}$ point at persistence value $d_i$. We calculate this value for all time windows, for each homologies $H_1$ to $H_4$, showing the average persistence as a function of time. This is done for topological features in increasing dimension. The results for each dimension are normalized by the average persistence of that homology over the entire time window. The resulting curve is shown in Fig.~\ref{fig:Finance_tda}. From this we immediate see that large peaks in the average persistence curve appear in all homologies in time windows close to extreme market crashes. These peaks are particularly noticeable near the 2008 financial crisis, the 2020 COVID-19 recession, and the 2022 stock market decline. 

This analysis both reproduces the signal obtained in previous work, and demonstrates that such a signal is also present in the higher order homologies. Notably, the relative size of the peaks produced by the higher order homologies visually appears to be larger than that produced by $H_1$ alone. Additionally, there are several large peaks that appear in the $H_3$ and $H_4$ persistence curves which are either absent or below the noise threshold in the $H_1$ curve. While there is some ambiguity in connecting these peaks to specific events, they do appear to correlate, for example to the lead up to the 2008 financial crisis, around the 2015 stock market sell-off and around the 2018 cryptocurrency crash.  

In order to quantify the degree and predictive nature of this correlation, we study the time-delayed Pearson correlation coefficient of the TDA signal with the S\&P 500 closing price. Let $X[t_0,t_f]$ and $Y[t_0,t_f]$ be the samples of the closing market price and the TDA signal in the interval between times $t_0$ and $t_f$. If we treat each set of sample as a random variable, we can then measure the covariance between the two distributions at a given time window, $T$. The average time-delayed correlation Pearson correlation is then given by
\begin{eqnarray}
    P(\tau) = \int \frac{\text{Cov}\big(X[t,t+T],Y[t-\tau,t+T-\tau]\big)}{\sigma_X\sigma_Y} dt.
\end{eqnarray}
That is, $P(\tau)$ measures the correlation between the two curves when the indicator (the TDA curve) is shifted by time $\tau$, relative to the signal (the market price).

\begin{figure}
    \centering
    
    \includegraphics[width=0.55\linewidth]{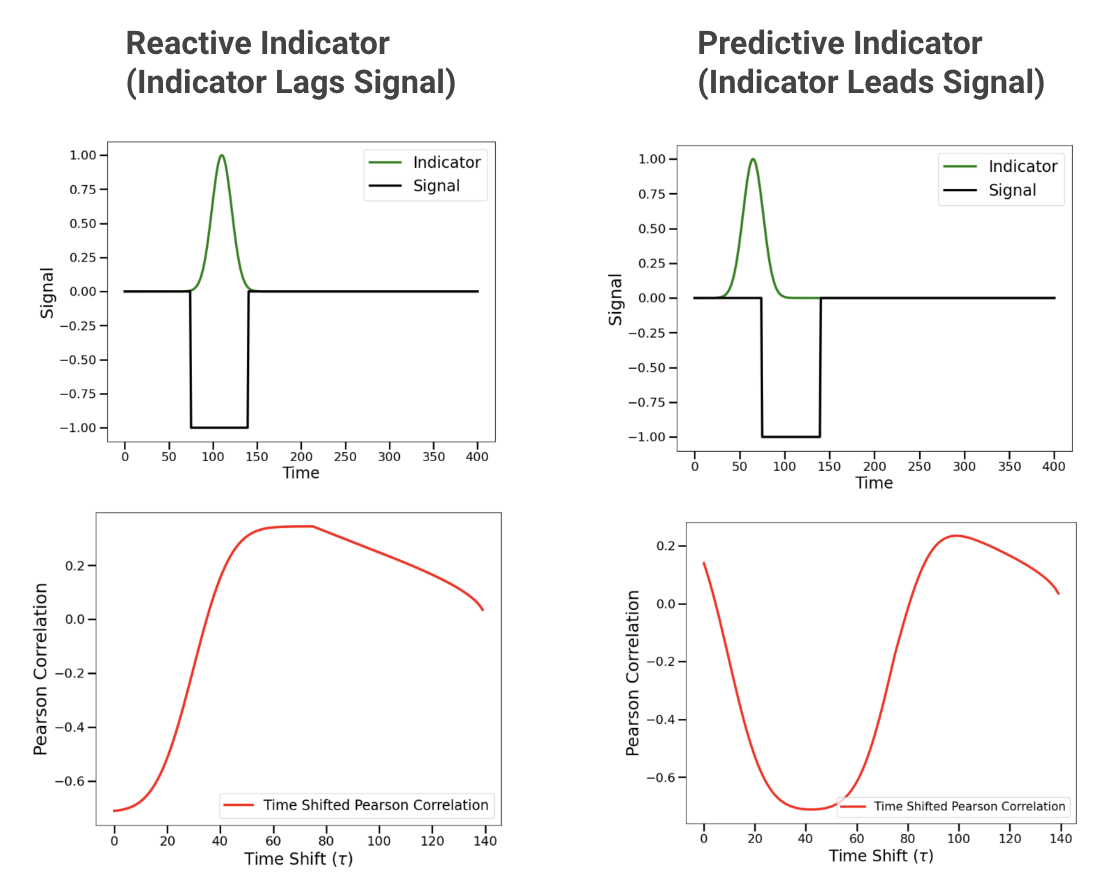}
    \caption{The TDA signal is an indicator which is negatively correlated with the stock price. By shifting the signal in time and measuring this correlation, we can measure the predictive potential of the TDA signal. In the cartoon example above, an indicator which lags the signal will still show a negative correlation with the price without shifting. However, as we shift the indicator signal to the future, the degree of correlation is reduced. On the other hand, an indicator which leads the signal and is therefore predictive will have a negative gradient in the time shifted correlation near time $\tau=0$.}
    \label{fig:time-delay_cartoon}
\end{figure}

In Fig.~\ref{fig:time-delay_cartoon}, we illustrate the general behavior of this Pearson correlation for a predictive indicator which leads the signal, vs that of a reactive indicator which only appears after the change in market price. As in our example, we assume that the indicator is negatively correlated with the signal. In this case, the behavior that we associate with a leading indicator is an initial drop in the correlation value with the time delay parameter $\tau$.

\begin{figure}
    \centering
    \includegraphics[width=0.46\linewidth]{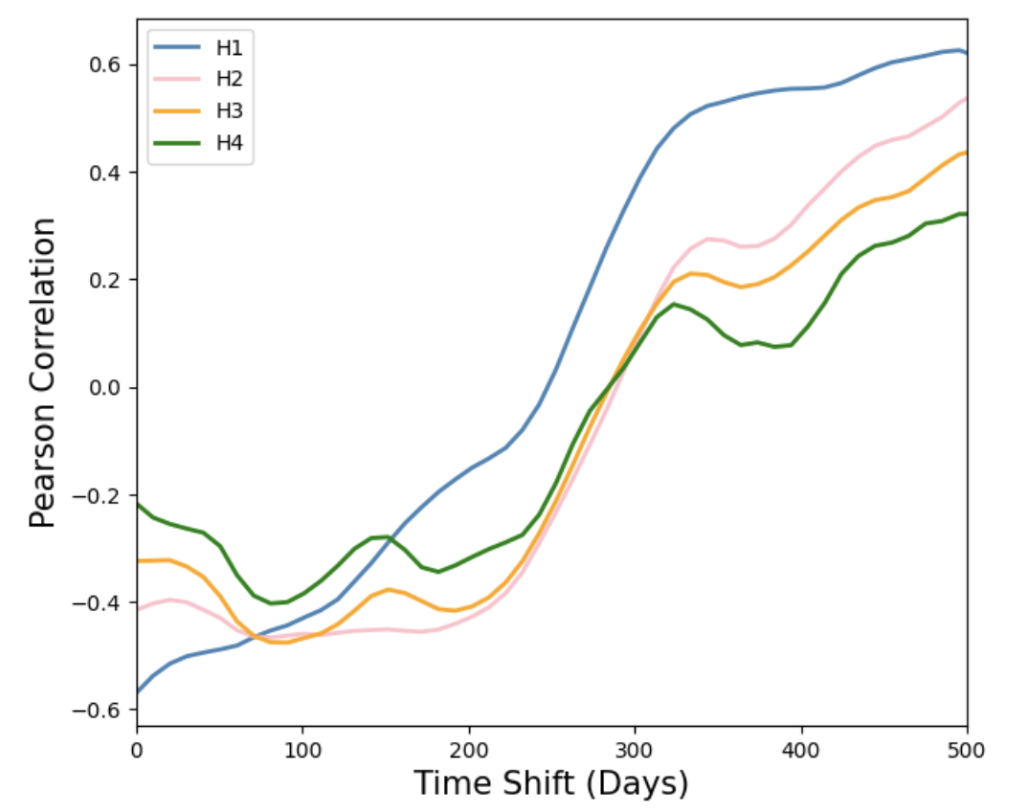}
    \quad\includegraphics[width=0.48\linewidth]{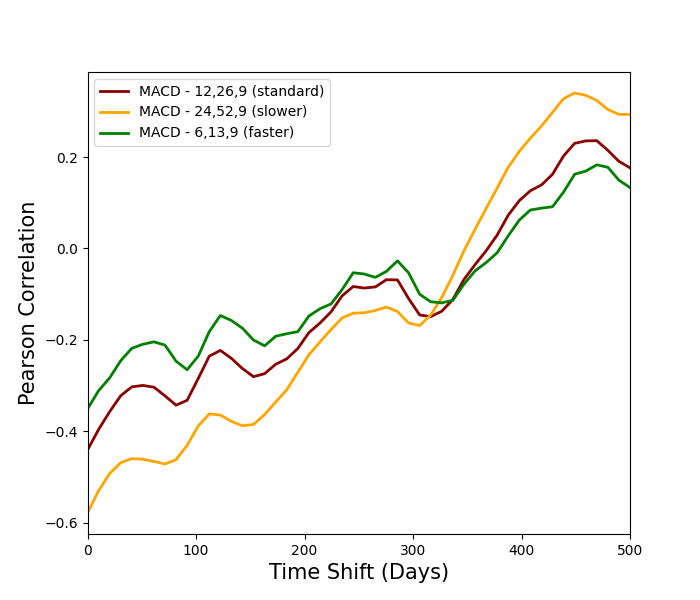}
    \caption{A comparison of the time-delayed Pearson correlations function of  TDA indicator signals for the first to fourth homology groups, compared to a simpler indicator, the MACD signal which is often used in finance applications. We see that the higher order homology curves show a stronger negative slope near $\tau=0$, indicating a higher degree of predictive power in predicting the market behavior.  }
    \label{fig:time-delay_data}
\end{figure}

In Fig.~\ref{fig:time-delay_data}, we measure the Pearson correlation function on the actual data, for a period of time between 2011 and 2018, where there are no large market events which skew the signal. We find that over this time period, the TDA signal from all homology groups $H_1$ to $H_4$ are negatively correlated with the overall market price. However, the higher order homologies show a stronger predictive nature as demonstrated by the initial decay followed by a maximum anti-correlation when shifted by $\tau\approx 100 days$. We compare this to another well known financial indicator, the moving average convergence/divergence (MACD). We compare to three different values over which we take the exponential moving averages for the MACD line and signal line. In all cases, we again find that the MACD signal is correlated with the closing price S\&P 500, however the MACD signal does not display the leading indicator behavior that the TDA signal shows.

\paragraph{Summary.}
In this section, we provided strong evidence that structural changes in point clouds constructed from financial time series data contains signals that can predict market downturns. Expanding on previous results in this area, we showed that the higher order homologies can provide a stronger signal in the vicinity of market crashes, and seems to detect several market downturns which did not give a strong signal in the lower order homologies. We also quantified the predictive power of the TDA signal by measuring the time-delayed Pearson correlation coefficient and find that the higher order homologies behave as an early predictive indicator, where the TDA signal precedes the onset of a market downturn. Overall, these results demonstrate that the information contained in the higher order topological structure of these financial point clouds provides new information which is hidden from traditional methods and is useful in understanding the collective behavior of a collection of assets over time. As we show in the following section, quantum TDA methods will allow us to extract even higher order topological features, which can uncover additional structure and give us further predictive power when analyzing this type of time series data.

\section{Quantum Topological Feature Extraction} \label{sec:qTDA_theory}

Here we assess the feasibility of using near-term quantum hardware to extract useful topological information from the combinatorial Laplacian, $\Delta_k^\Gamma$. Rather than attempting to compute full persistent homology or exact Betti numbers in the most general setting, we focus on a more application-driven task: estimating low-order spectral moments of $\Delta_k^\Gamma$ that can serve as a proxy for high-dimensional topological structure. We present evidence that these low-order features can be strongly correlated with high-dimensional Betti information in the relevant regimes. This reframing shifts the near-term goal from exact computation of Betti numbers to extraction of topological features that are useful for downstream analysis.  We demonstrate this approach on a barium-based trapped-ion development system, extracting Laplacian-derived features from graphs with up to 16 nodes. To our knowledge, these experiments constitute the first quantum-hardware demonstration of moment-based quantum TDA in which a measured Laplacian observable is quantitatively compared against exact Betti information to distinguish graph instances with different high-dimensional topology.

We present a circuit construction and resource analysis for this task, including estimates of the qubit count, two-qubit gate count, and shot requirements needed to measure these observables on current and near-term hardware. Our analysis identifies regimes of simplicial complexes in which the cost of extracting these features scales far more favorably than one might expect from conventional Betti number-estimation arguments. In particular, the usefulness of the method does not appear to require the relative Betti number $\beta_{k-1}/N_k$ to be large. Taken together, these results suggest a concrete and experimentally testable pathway by which quantum TDA could provide practical value before fully fault-tolerant quantum computers are available.



\subsection{Resource Requirements for Moment-Based Quantum TDA} \label{sec:resources}

Here we outline the general resources required to implement the NISQ Quantum TDA (QTDA) algorithm. The goal of the quantum algorithm is to efficiently extract properties of the combinatorial Laplacian that encode topological information about a given simplicial complex.  We will describe the different gadgets that can be used to efficiently construct this combinatorial Laplacian operator and analyze the methods that can be used to implement these gadgets. This will allow us to give accurate estimates of the number of qubits, two-qubit gates, and shot counts needed to extract the relevant properties from this operator. We then discuss the feasibility of near and intermediate term implementations of the QTDA algorithm on trapped ion quantum computers.

Broadly, there are three components in the quantum TDA algorithm: data loading of the simplicial complex, application of the Laplacian operator, $\Delta^\Gamma_k$, and quantum trace estimation. The data loading involves preparing a quantum state that is a superposition over all k-cliques of the given simplicial complex. The application of the combinatorial Laplacian operator involves acting on this initial state with a simple fermionic Hamiltonian that is projected into the space of the simplicial complex. For a simplicial complex with $N$ nodes, and $|E|$ edges, we can construct quantum circuits for both of these components using $\mathcal{O}(N)$ qubits and $\mathcal{O}\bigg(N+\min(|E|,\binom{N}{2}-|E|)\bigg)$ CNOT gates. 

We then aim to measure the relative trace of moments of the Laplacian, $\operatorname{tr}[(\Delta_k^\Gamma)^p]$. Throughout this section, we use $\operatorname{Tr}$ to denote the ordinary trace and define the normalized trace over the $k$-clique subspace by 
\begin{eqnarray}
    \operatorname{tr}[\Delta_k^\Gamma] = \frac{1}{N_k}\operatorname{Tr}[\Delta_k^\Gamma].
\end{eqnarray}
For most of this section, we will in fact focus on measuring the relative trace of only the first moment, $p=1$. This can be done by averaging the expectation values $\bra{\ell}\Delta^\Gamma_k \ket{\ell}$ over random initial wavefunctions $\ket{\ell}$ whose basis states contain different random phases factors. Both the data loading state preparation and the trace estimation procedures require repeated execution of the quantum circuit. As we will see below, for the data loading, a repeat until success procedure is used that on average requires $\zeta_k^{-1} \sim \mathcal{O}(\zeta_2^{-ak^b})$ attempts to succeed, for some power $b$, which is independent of $N$. For  trace estimation, we are able to measure the normalized trace to within additive accuracy $\epsilon$ using $\mathcal{O}(1/\epsilon^2)$ random initial vectors $\ket{\ell}$, each of which is sampled a constant $M$ number of times. As we will see in Sec.~\ref{sec:rel_tr}, we only require a fixed precision $\epsilon$ that is independent of $N$ and $k$, to extract meaningful information about the topological structure of the complex. Also, in contrast to the common belief surrounding quantum TDA methods, our analysis suggests that the quantum complexity of extracting the topological information from the simplicial complex depends only very weakly on the value of the relative Betti number $\beta_{k-1}/N_k$, where $N_k$ is the number of k-cliques. Therefore, the overall complexity of our algorithm scales as 

\begin{eqnarray}
\mathcal{O}\bigg(\bigg[N+ \binom{N}{2}\min(\zeta_2,1-\zeta_2) \bigg]\,\zeta_2^{-a k^b}/\epsilon^2 \bigg),
\end{eqnarray}
where $\zeta_2 = |E|/\binom{N}{2}$.

\subsubsection{Circuit Width}

We encode a simplicial complex in quantum states where each qubit represents a single graph node. To study the degree-$d$ Betti number $\beta_d$, we work in the Hamming-weight $k=d+1$ subspace. These states represent $k$-cliques, equivalently $d$-simplices. Therefore, the quantum algorithm requires at least $N$ qubits to encode a simplicial complex on $N$ nodes. As described in Ref.~\cite{akhalwaya2023topological}, one may use additional ancilla qubits to implement the different components of the quantum algorithm. In general, the number of ancilla qubits is a design choice that can be varied to trade circuit width for depth and parallelization of measurements. In our circuit construction, we use ancilla qubits both for the Dicke state preparation module and for the projection onto the complex $\Gamma$.  

 The projection operator applies a Toffoli gate from each edge in the complement graph $\bar{\Gamma}$ to an ancilla qubit, which is then measured and reset. In principle, only a single ancilla qubit is needed for this step; however, in this case, we would need to perform mid-circuit measurements after every CCX gate, requiring $(1-\zeta_2)N^2$ separate measurement rounds. On the other extreme, no mid-circuit measurement is needed if we use $(1-\zeta_2)N^2$ ancilla qubits. The application of the Toffoli gates can be maximally parallelized by applying the CCX gates on disjoint sets of edges in parallel and measuring all qubits once. This would require $N/2$ ancilla qubits and would maximally reduce the number of circuit layers needed for data loading. Using this scheme gives
\begin{eqnarray}
    \text{\# of qubits} = \frac{3}{2}N .
\end{eqnarray}

\subsubsection{Circuit Depth}

Here we describe the construction of a quantum circuit that applies the first moment of the combinatorial Laplacian. This operator can be written as $\Delta^\Gamma_k = P_k \Delta^\Gamma P_k = P_k P_\Gamma B P_\Gamma B P_\Gamma P_k$, where $P_k$ and $P_\Gamma$ are projectors onto the set of k-cliques and onto the simplicial complex $\Gamma$, respectively. The boundary operator $B = \partial + \partial^\dagger$ maps k-cliques to $k\pm 1$ cliques, with an appropriate sign indicating its direction.  In the language of quantum circuits, this operator can be implemented using three major gadgets.

The first gadget constructs a Hamming weight k Dicke state on $N$ qubits, $\ket{D_k^N}$, with random phases between the basis states in order to average over initial vectors $\ell$. The Dicke state preparation can be accomplished in several ways \cite{childs2000finding,bartschi2022short,bartschi2019deterministic}. The most efficient in terms of gate depth probabilistically prepares $\ket{D_k^N}$ by applying a quantum phase routine with Hamiltonian $H = \frac{1}{2}\sum_i (1-Z_i)$ to project a suitable initial product state into a fixed Hamming-weight wavefunction. The initial product state is chosen to have a Hamming weight distribution peaked around $k$. Since $H$ contains only single qubit gates, the full QPE routine requires only $2\lceil \log_2(k)\rceil N$ CX gates. The final state $\ket{D_k^N}$ represents a sum over all possible k-cliques in the simplicial complex on $N$ nodes

The second gadget then projects this Dicke state onto only basis states representing k-cliques that are in the complex $\Gamma$. Let $\zeta_k$ be the density of k-cliques in the complex, $\zeta_k = N_k/\binom{N}{k}$. Note that $\zeta_2$ is the density of edges, $\zeta_2 =|E|/\binom{N}{2}$. We focus here on the case where $\zeta_2 > 0.5$, since that is the most likely use case for quantum TDA applications. Here, we follow the procedure outlined in \cite{akhalwaya2023topological}, whereby we may project the $\ket{D_k^N}$ Dicke state onto a specific k-clique complex by acting one Toffoli gate for each edge in the complement graph $\bar{\Gamma}$. That is, we must apply a number of gates directly proportional to $(1-\zeta_2)$, the number of edges \emph{missing} from the graph.  In particular, we can apply a Toffoli gate with controls on the qubits representing the vertices of a missing edge, to a target ancilla qubit, $U_e = CCX_{i,j,a}$ for $(i,j) \in \bar{E}$. When acting $\prod_{e \in \bar{E}} U_e$ on a Hamming weight k basis state $\ket{x}\ket{a}$, the ancilla $\ket{a}$ will be flipped iff $\ket{x}$ does not represent a k-clique of $\Gamma$, since in that case $\ket{x}$ contains both vertices of an edge in $\bar{E}$. The algorithm therefore proceeds by applying $\prod_{e \in \bar{E}} U_e$, with one target ancilla per $U_e$ operator. Measuring all ancilla qubits in the $\ket{0}$ state results in the preparation of the desired target state $\psi = \sum_{x \in \Gamma_k}\ket{x}$. As explained in the previous sections, we may use mid-circuit measurement and reset operations to avoid the need for $\mathcal{O}(N^2)$ ancilla qubits. The procedure succeeds with a probability proportional to $\zeta_k$. We discuss the behavior of this success probability in the next subsection. 

The final gadget implements the unconstrained boundary operator $B = \partial + \partial^\dagger$. In the quantum mapping, this operator is equivalent to the free fermion operator $B = \sum_i (a_i + a_i^\dagger)$, which adds and removes vertices in the Hamming weight-k Dicke states with a relative sign equivalent to that of the boundary operator from algebraic topology. The quantum operator $B$, can be written as a sum of Jordan-Wigner strings, 
\begin{eqnarray}
    B = \sum_i \left[\prod_{j<i} Z_j\right]X_i.
\end{eqnarray}
Note that this operator is both a sum of unitaries and $B/\sqrt{N}$ is itself unitary. We can therefore implement the operator 
\begin{eqnarray}
    U = e^{i\theta B} = \cos(\theta\sqrt{N})I + i \sin(\theta\sqrt{N}) B/\sqrt{N} \label{eq:eiB}
\end{eqnarray} 
 using Trotterization methods.  Each Trotter step  can be implemented with $4N-2$ CX gates. We apply $T$ trotter steps, with $T$ scaling as $\mathcal{O}(\sqrt{N})$.  

Putting these three gadgets together allows us to apply the circuit
\begin{eqnarray}
    \ket{\psi_\ell} =  P_\Gamma B P_\Gamma P_k |\ell\rangle = P_\Gamma B P_\Gamma \ket{D_k^N}_\ell,
\end{eqnarray}
where $\ket{D_k^N}_\ell$ is a Dicke state on N qubits with random phases assigned to between the Hamming weight k basis states. Since we want to randomize over these phases anyway, the first $P_\Gamma$ operator is also allowed to apply random phase operations. This allows us to use the relative implementation of the Toffoli gates in the construction of $P_\Gamma$, where each Toffoli gate is compiled to only 3 CX gates. In order to measure $\text{tr}[\Delta^\Gamma_k] = \frac{1}{M}\sum_\ell \phantom{.}_\ell\langle D_k^N| P_\Gamma B P_\Gamma B P_\Gamma \ket{D_k^N}_\ell$, we estimate the norm of the wavefunction $||\ket{\Psi}_\ell||^2$. This allows us to apply the second $P_\Gamma$ operator using only classical post-selection on the measured bit-strings. 

Putting this together, we find that the overall scaling of CNOT gate count is given by
\begin{eqnarray}
    \text{\# of CNOTS} &=&  3(1-\zeta_2)\frac{N(N-1)}{2} \,+\, \bigg [ 2N\lceil\log_2(k+1)\rceil \nonumber \\ &+& \frac{\lceil\log_2(k+1)\rceil(\lceil\log_2(k+1)\rceil-1)}{2} \bigg] \,+\, T(4N-2) 
\end{eqnarray}
where $\zeta_2 = |E|/(N(N-1)/2)$, is the ratio of the number of edges in the graph to the maximum number of edges, and T is the Trotter order.

\subsubsection{Shot Counts}

The number of shots needed to measure the relative trace, $\operatorname{tr}[\Delta^\Gamma_k]$, depends mainly on the desired accuracy $\epsilon$ and the success probability of the data loading procedure that depends on projecting into the simplicial complex $\Gamma$. In order to estimate
\begin{eqnarray}
    \operatorname{tr}[\Delta_k^\Gamma] &\approx& \frac{1}{M}\sum_{\ell=0}^M \langle v_\ell| \Delta_k^\Gamma |v_\ell\rangle \\
    &=& \frac{1}{M}\sum_\ell \langle v_\ell| P_k P_\Gamma B P_\Gamma B P_\Gamma P_k |v_\ell\rangle, \label{eq:rel_trace}
\end{eqnarray}
we create the state
\begin{eqnarray}
    \ket{\Psi}_\ell =  P_\Gamma B P_\Gamma P_k |v_\ell\rangle,
\end{eqnarray}
where $\ket{v_\ell}$ is a random Hadamard vector (a sum over all basis states $\ket{x}$ with random signs on each state $\ket{x}$). We then calculate the norm $||\ket{\Psi}_\ell||$ through repeated measurements of all qubits.

The number of different vectors, $M$, that are used determines the final precision, $\epsilon$, on the estimate $\operatorname{tr}[\Delta_k^\Gamma]$. For a given $\epsilon$, we need to choose $M \sim 1/\epsilon^2$. Experimentally, for Erd\"os-R\'enyi random graphs, we find that
choosing $M\sim 100$ gives $\epsilon \sim 0.1$, which is approximately the accuracy needed to distinguish graphs with high vs low Betti number based on the results of the next section.  
The second factor that determines the total shot count is how many shots are necessary for \emph{each} vector $\ket{v_\ell}$, to estimate $E_\ell = \langle v_\ell|\Delta_k ^\Gamma\ket{v_\ell}$. Our overall error target for the average is given by $\epsilon$. This implies the maximum tolerable error for each individual expectation value $E_\ell$ is $\sqrt{M}\epsilon$ (using standard propagation of uncertainty formulas, the error of the sum $\sum_{\ell=0}^M E_\ell$ will be $\epsilon_\ell/\sqrt{M}$). 

The main difficulty now comes in determining how many shots $M_{\ell,\Gamma}$ make it through the first projection operator $P_\Gamma$. That is, if we run $M_\ell$ circuits, what is the value $\langle v_\ell|P_\Gamma P_k |v_\ell\rangle$. This is the number of ``usable'' shots out of the $M_\ell$ initial shots. This value depends on the value of 
\begin{eqnarray}
    \zeta_k = \frac{\text{\# of k-cliques in $\Gamma$}}{\binom {N}{k}}.
\end{eqnarray}

Therefore, the total number of shots per vector is given by
\begin{eqnarray}
    M_\ell = \frac{1}{\zeta_k} \sigma,
\end{eqnarray}
where $\sigma$ is the standard deviation of the expectation value for a single vector, and may depend on the specific simplicial complex.

\begin{figure}[H]
    \centering
    \includegraphics[width=.95\linewidth]{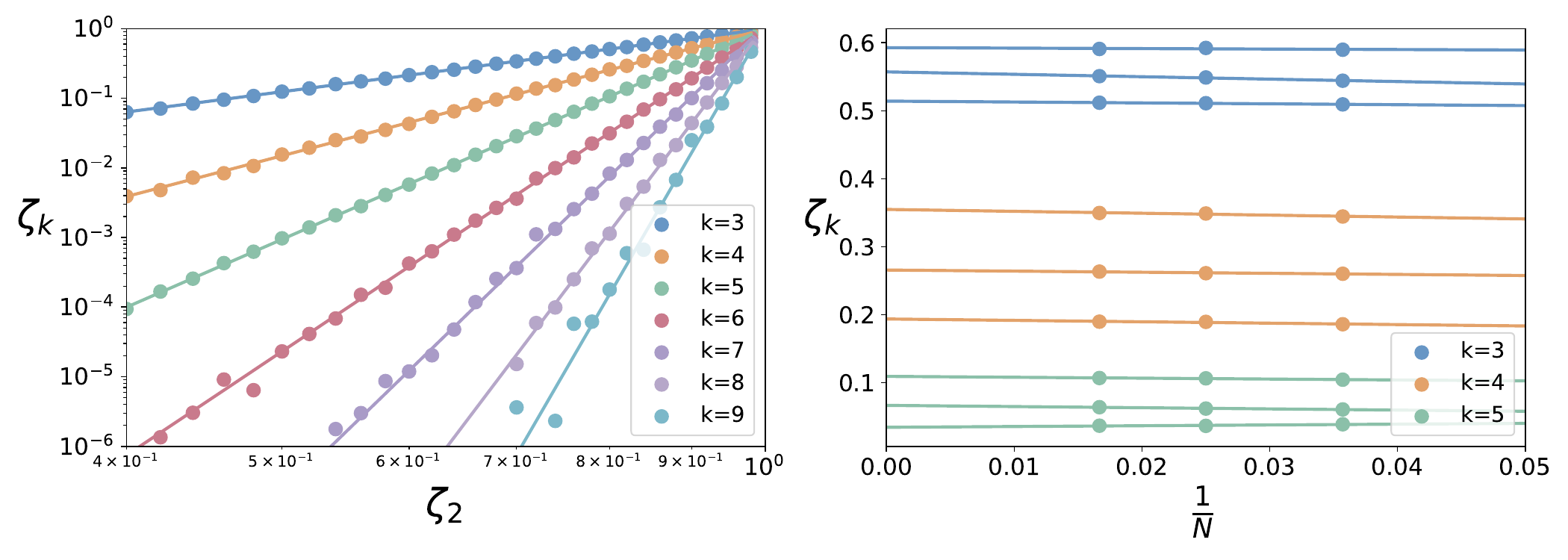}
    \caption{{\it (Left)} Relationship between $\zeta_2$ and $\zeta_k$ for random graphs. For ensembles of Erd\"os-R\'enyi graphs, we find that $\zeta_k = \zeta_2^{a\cdot k^b}$ with $a\approx 0.23(1)$ and $b\approx 2.3(5)$, so that the ratio $\zeta_k$ rapidly decreases for higher clique complexes at fixed edge density $\zeta_2$. {\it (Right)} We show the scaling of $\zeta_k$ with $\frac{1}{N}$ at three different fixed values of $\zeta_2$, ($\zeta_2=0.82,0.78,0.74$) for three different values of $k$. A nearly flat scaling with $N$ over $N=28,40,60$ is seen across all parameter regimes, indicating an independence of scaling of the required shots with $N$ at fixed $k$.}
    \label{fig:zeta2_vs_zetak}
\end{figure}

We can see the relationship between $\zeta_2$ and $\zeta_k$ in random graphs. This relationship is plotted in Fig.~\ref{fig:zeta2_vs_zetak}.

From this relationship, we see that at \emph{fixed} $\zeta_2$, the value of $\zeta_k$ decreases exponentially, roughly as $1/2^k$. In this case, for the same graph, we need $2^k$ times as many shots as $k$ increases. Notably, for fixed $k$, this value is independent of the size of the graph $N$. That is, we do not require more shots as the size of the graph increases. 

However, we can always increase $\zeta_k$ by looking at graphs that are more dense. In the extreme case, we choose a value for $\zeta_k$, and essentially only consider graphs such that $\zeta_k > cutoff$. So that the total number of shots does not increase as we increase $k$, but the number of graphs that we can study decreases as $k$ increases. 

Finally, we note that we can decrease the number of shots needed for each $\ket{v_\ell}$ by performing amplitude amplification on the state $P_\Gamma \ket{k_\ell}$. This amplitude amplification step involves applying $2x$ additional repetitions of the $P_\Gamma$ operator, and increases the probability of measuring a state in the good subspace by a factor of $(2x+1)^2$. If we target $k=8-12$, and $\zeta_2=0.8$, it may be beneficial to use this with $x=1-3$.

While at low values of $\zeta_2$, we find that $\zeta_k$ becomes extremely small ($<10^{-10}$), implying that the number of shots required becomes intractable in this regime, the situation is much more promising at higher edge density. For $\zeta_2=0.85$, we find that $\zeta_k$ varies between $\zeta_3=0.65$ and $\zeta_{10}=0.0005$, indicating that even relatively high values of $k$ are within reach of quantum algorithms.  On top of this, for higher values of $k$, we expect that the existence of nontrivial Betti numbers likely requires high edge density, coinciding with the regime where the quantum algorithm is tractable.



\subsection{Effectiveness of the Relative Trace as a Topological Feature}\label{sec:rel_tr}

In the quantum TDA algorithm, we seek to learn the degree-$d$ topological features of a given simplicial complex by analyzing the eigenvalue spectrum of the associated combinatorial Laplacian. The conventional wisdom assumes that our ability to estimate the Betti number depends strongly on the ratio of the Betti number to the total number of cliques in the complex \cite{babbush2025grand}.  In contrast to this, we show that a heuristic approach which relies instead on learning a relationship between the relative trace of the Laplacian and the Betti number allows us to approximately extract $\beta_d$ even in instances when the relative Betti number $\beta_{k-1}/N_k$ is very small. This greatly increases the range of situations when quantum TDA may be effective in the near term. 

In order to estimate the Betti number, the typical approach proposed in the literature is to measure the trace m$^\text{th}$ moment of $\Delta^\Gamma_k$, and estimate the number of zero eigenvalues of this operator using a Chebyshev expansion of the Heaviside function $H(\delta I- \Delta^\Gamma_k)$, so that
\begin{eqnarray}
   \beta_{k-1} = \text{Tr}[H(\delta I- \Delta^\Gamma_k)] &=& \sum_i H(\delta-\lambda_i) |\lambda_i\rangle\langle\lambda_i| \\
    &=& \sum_m c_m \text{Tr}\bigg[\big(\Delta^\Gamma_k\big)^m\bigg],
\end{eqnarray}
$\lambda_i$ are the eigenvalues of $\Delta^\Gamma_k$ and $\delta$ is the estimated energy gap to the first nonzero eigenvalue. Since we know that $\text{Tr}[(\Delta^\Gamma_k)]^m = \sum_i \lambda_i^m$, we see that we are approximately counting the number of zero eigenvalues of the matrix $\Delta^\Gamma_k$, by estimating the sum of powers of the eigenvalues $\text{Tr}[(\Delta^\Gamma_k)^m] = \sum_i \lambda_i^m$. The conventional wisdom then assumes that our ability to estimate the Betti number $\beta_{k-1}$, to precision $\epsilon$ depends strongly on the ratio of $\beta_{k-1}$ to the total number of cliques $N_k$.  That is, the effect of the number of zero eigenvalues on this sum depends on the ratio of the number of such zero eigenvalues, (i.e. the Betti number $\beta_{k-1}$) to the total number of eigenvalues of $\Delta^\Gamma_k$ (i.e. the number of k-cliques in $\Gamma$). In the quantum algorithm, this is manifest in the need for a precision of $\epsilon \lesssim \beta_{k-1} / N_k$, when measuring the relative trace observable, $\text{Tr}[\Delta^\Gamma_k] =  \sum_\ell \bra{\ell}\Delta^\Gamma_k \ket{\ell}$.  This seems to imply that we must either 1) measure the observable to very high precision, or 2) limit our study to simplicial complexes where $\beta_{k-1}/N_k$ is large. In both cases, this can severely impact the general ability of the quantum TDA algorithm to impact real world applications. 

In contrast to this, we will now show that for a wide family of graphs, there exists a strong correlation between the first moment of the relative trace $\text{Tr}[\Delta^\Gamma_k]$, and the Betti number, even in cases when the relative Betti number, $\beta_{k-1}/N_k$, is very small. In fact, amazingly, in many cases the degree of correlation appears to be nearly independent of the relative Betti number completely. This correlation can allow one to perform a linear regression to estimate the Betti number from the quantum relative trace without the need to perform the exact Betti number calculation. We numerically demonstrate this relationship using random Erd\H{o}s-R\'enyi graphs.  We generate ensembles of random graphs with the exact same number of edges, and approximately the same number of k-cliques. We then measure the correlation between the relative trace and the degree-$d=k-1$ Betti number $\beta_{k-1}$, both of which are exactly calculated classically. 

\begin{figure}
    \centering
    \includegraphics[width=0.98\linewidth]{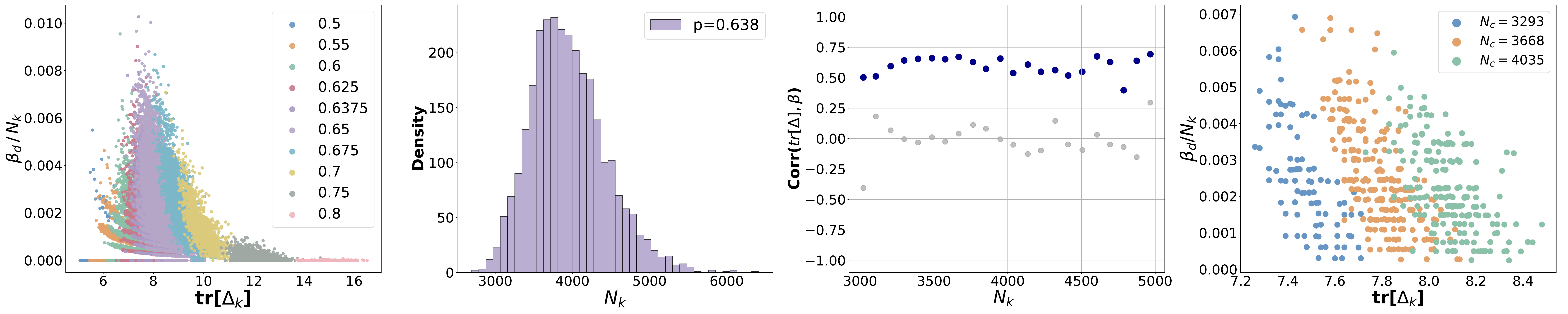}
    \includegraphics[width=0.98\linewidth]{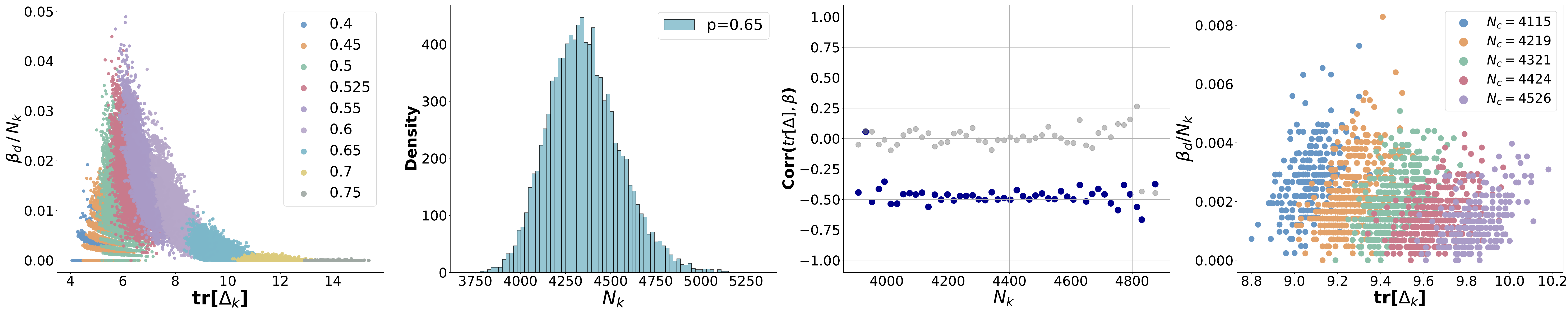}
    \caption{Distributions of graphs demonstrating correlation between the Betti number $\beta_d$ and the relative trace $\text{Tr}[\Delta^\Gamma_k]/N_k$ for graphs with $N=36$ nodes and $k=5$ {\it (top row)} and $k=4$ {\it (bottom row)}. From left to right we show the distribution of $\beta_{k-1}$ and RT over graphs with the same edge density. The second plot shows the distribution of $N_k$ values within an ensemble at fixed $\zeta_2$.  The third plot shows the correlation between the $\beta_k$ and RT within each bin of graphs with similar $N_k$. The far right plot shows samples of the distributions within specific bins. }
    \label{fig:betti_trace_corr}
\end{figure}

For random Erd\H{o}s-R\'enyi graphs, a demonstration of these are shown in Fig.~\ref{fig:betti_trace_corr}. We plot the distribution of $\beta_{k-1}$ vs the relative trace for ensembles of graphs with different edge densities, $\zeta_2$. For each value of $\zeta_2$, there is a varying degree of correlation between these values, however the strength of correlation is obscured by the fact that both the Betti number and the relative trace are correlated with the density of k-cliques,  $\zeta_k$, which can vary significantly within graphs with the same $\zeta_2$. To compensate for this, we bin the ensemble into groups with similar values of $\zeta_k$, and measure the correlation within each bin. After this binning procedure, we see a much stronger degree of correlation between $\text{Tr}[\Delta^\Gamma_k]$ and $\beta_{k-1}$. Interestingly, the sign of this correlation appears to depend on the sign of the derivative $\partial \beta_k /\partial(Tr\Delta) $.  This is in contrast to the \emph{naive} expectation that a larger value of $\beta/N_k$ implies a smaller value of the relative trace since a significant portion of the eigenspectrum is filled with zeros.  For high values of $\zeta_2$, we see a positive correlation emerges between the two parameters, despite the fact that there is a negative correlation in the full distribution when not binned by $\zeta_k$, a demonstration of Simpson's paradox.  

\begin{figure}
    \centering
    \includegraphics[width=0.9\linewidth]{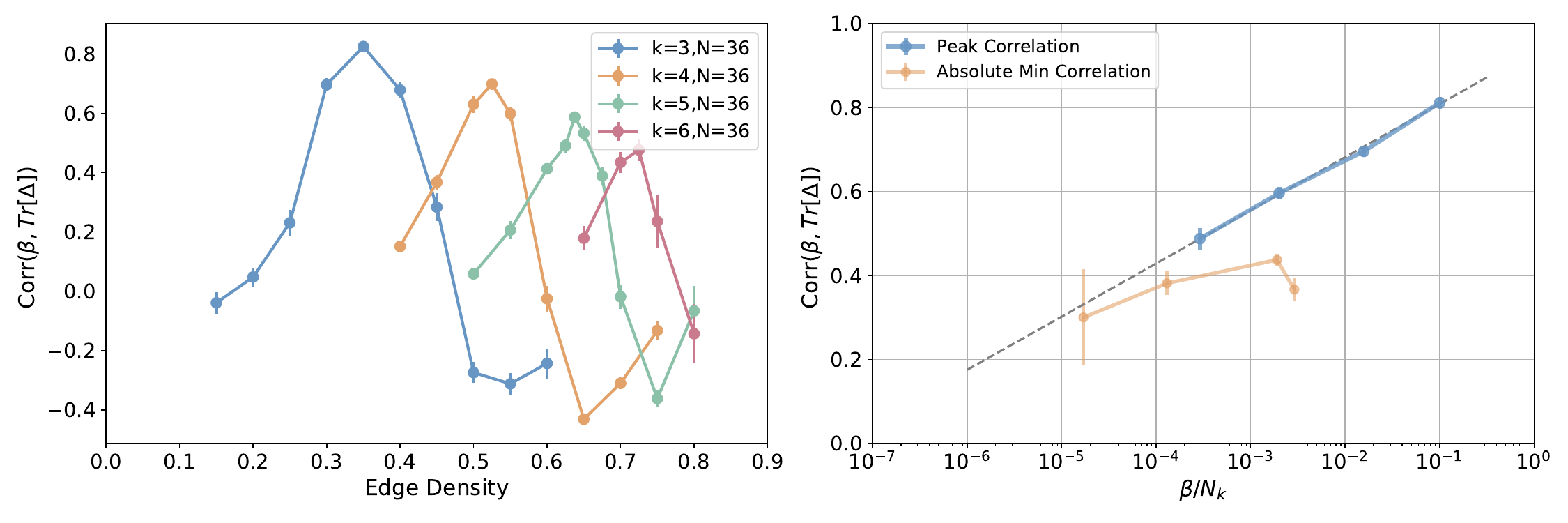}\label{fig:erdos_renyi_corr}
    \caption{{\it (Left)} Average correlation between the k Betti number $\beta_k$ and the relative trace $\text{Tr}[\Delta^\Gamma_k]/N_k$, as a function of edge density for k=3,4,5,6 over an ensemble of random graphs on $N=36$ nodes. For each edge density, graphs are binned into 40 groups according to the clique density $N_k$, and the correlation is averaged over all bins which contain at least 20 graphs. {(\it Right)} The peak of the correlation curves as a function of the relative Betti value $\beta_k/N_k$. We see that the degree of correlation within the ensembles ,decays only logarithmically with the relative trace at the peak value, and possibly slower at the anti-peak.}
\end{figure}

\subsubsection{Correlation on fMRI Point Clouds}
\begin{figure}[H]
    \centering
    {\includegraphics[width=0.35\linewidth]{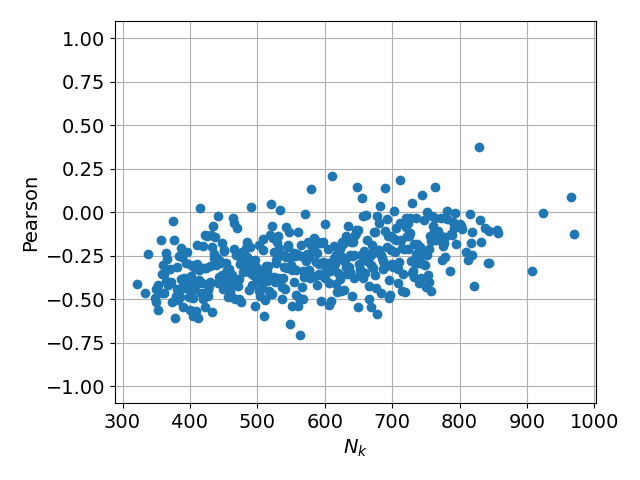}} 
    {\includegraphics[width=0.35\linewidth]{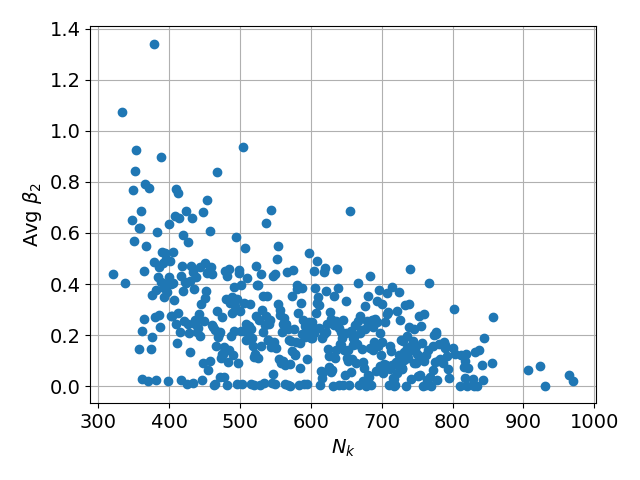}} \\
    
    {\includegraphics[width=0.35\linewidth]{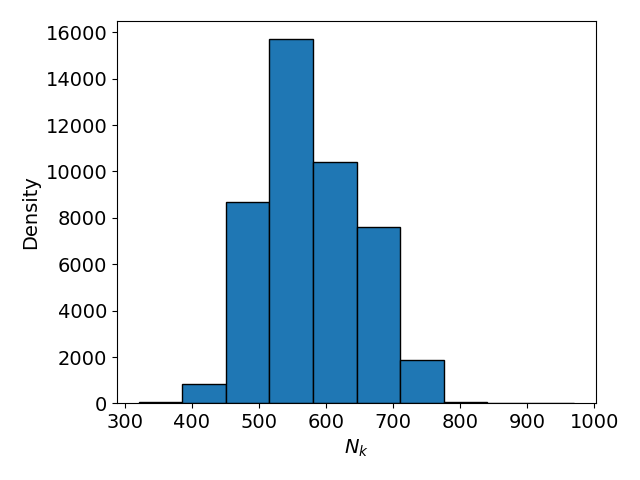}}
    {\includegraphics[width=0.35\linewidth]{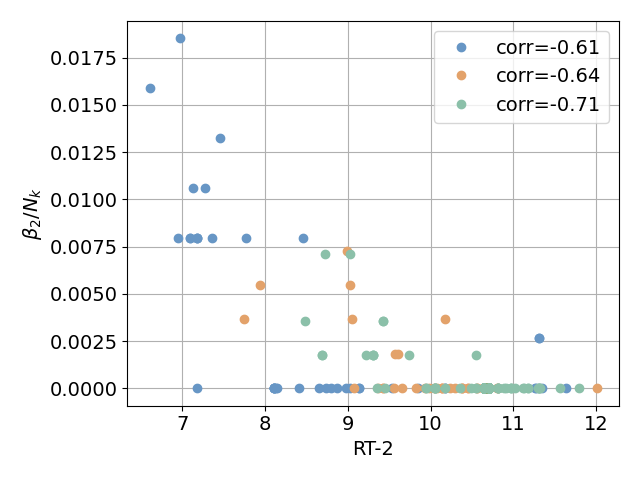}}
    \caption{To extract similar characteristics between Betti and relative trace as in Fig.~\ref{fig:betti_trace_corr}, we use the Alzheimer's fMRI point clouds of dimension $65$ and $26$ points at filtration values where edge density $p=0.6$. This is also the edge density where strongest correlation between $\beta_2$ and relative trace-$2$ was seen. {\it (top left)} Correlation between $\beta_2$ and $\text{Tr}[\Delta^\Gamma_k]/N_k$ with number of $k=3$ cliques, $N_2$, for all graphs with edge density $p=0.6$. {\it (top right)} average $\beta_2$ over all graphs in each $N_2$. {\it (bottom left)} density of graphs over all $N_2$ bins within $p=0.6$. {\it (bottom right)} Relative $\beta_2$ vs $\text{Tr}[\Delta^\Gamma_k]/N_k$ for $N_k$s with the highest correlation values.}
    \label{fig:hnorm_vs_RT_dim65}
\end{figure}

In this section, we demonstrate correlation between the Betti numbers and the expected output from the quantum algorithm (which is the relative trace ) on the real-world fMRI dataset we used in Sec.~\ref{sec:brainfmri}.In Fig.~\ref{fig:hnorm_vs_RT_dim65} we compress each fMRI time series to a very high-dimensional point cloud with $65$ dimensions and $\sim 26$ points. Each ROI from a patient gives one such high-dimensional point cloud. The input to the quantum circuit are graphs from the point clouds at specific filtration values. Similar to Fig.~\ref{fig:betti_trace_corr}, the input graphs are chosen at certain edge densities where $\beta$ and relative trace have strong correlations, and then the graphs are grouped into bins with the same number of k-cliques $N_k$. 
We demonstrate in Fig.~\ref{fig:hnorm_vs_RT_dim65} that the relative Betti ($\beta_{k-1}/N_k$) is high at low values of $N_k$, which is consistent with the strong correlation shown by $\beta_{k-1}$ and relative trace at those values.




\subsection{Quantum-Classical Crossover Estimates}

The resource analysis presented in Sec.~\ref{sec:resources} can be used to estimate the time-to-solution (TTS) of performing Betti number estimation using the QTDA algorithm. The relevant parameters for the TTS calculation are the number of graph nodes, N, the edge density, $\zeta_2$, the homological degree $d$ or equivalently the clique size $k=d+1$, and the desired precision $\epsilon$. The total TTS can be estimated using reasonable parameters for the speed of the quantum gate and measurement operations on the quantum hardware. This can be compared to the time-to-solution estimates of the classical TDA software presented in Sec.~\ref{sec:classical_alg}, which depends only on the number of nodes and homological dimension.  Before proceeding, we must acknowledge that this is not a completely fair comparison, as the classical TDA algorithm gives an exact value for the Betti number across a full filtration while our QTDA algorithm at low order in powers of $(\Delta_k^\Gamma)$ can only estimate an observable which is correlated to exact Betti number. Indeed, in the results we present below we look only at the first power of $\Delta_k^\Gamma$, which we expect can efficiently be estimated with alternative classical approaches. On the other hand, we expect that slight adjustments to the quantum algorithm will be able to improve on the Betti number estimation algorithm while retaining the same order of magnitude execution time and will be difficult to reproduce efficiently with any classical algorithm.  Therefore, the crossover estimates provided should be seen in the context of providing an approximate comparison a family of new quantum algorithm to the baseline classical approach used by current TDA practitioners. 

We use mildly optimistic estimates for the two-qubit gate speeds of 10$\mu s$ per gate, which is approximately an order of magnitude faster than current trapped ion hardware speeds, but within the range of speeds that have been demonstrated in specialized trapped ion experiments \cite{savill2026high,srinivas2021high}. On the other hand, we also assume serialized gate operations, consistent with current generation quantum hardware. We note that effective execution time of the quantum algorithm can realistically be improved via the use of parallel gates and algorithmic improvements such as the application of quantum amplitude amplification subroutines. 

\begin{figure}[H]
    \centering
    \includegraphics[width=0.4\linewidth]{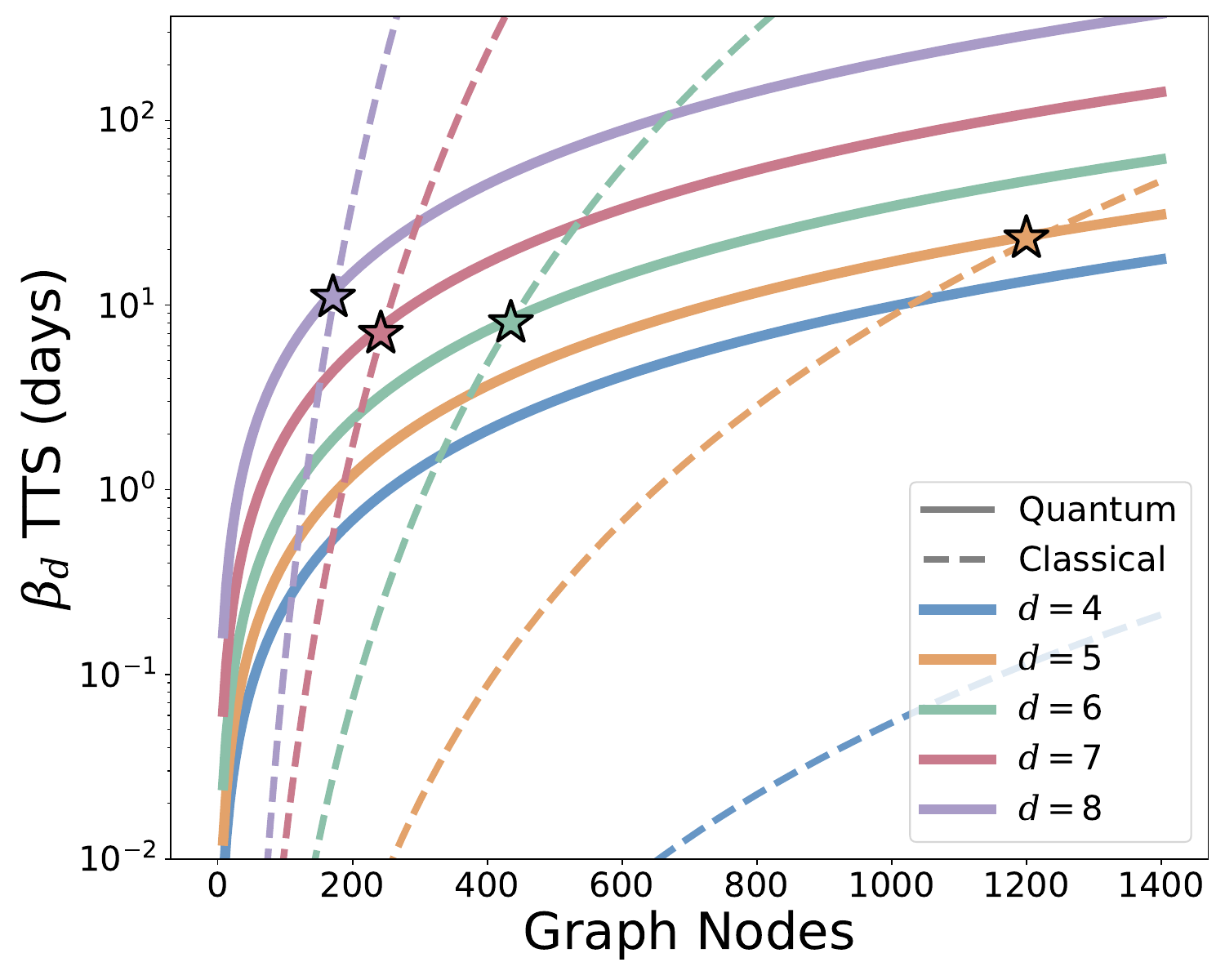}
    \hspace{10mm} \includegraphics[width=0.4\linewidth]{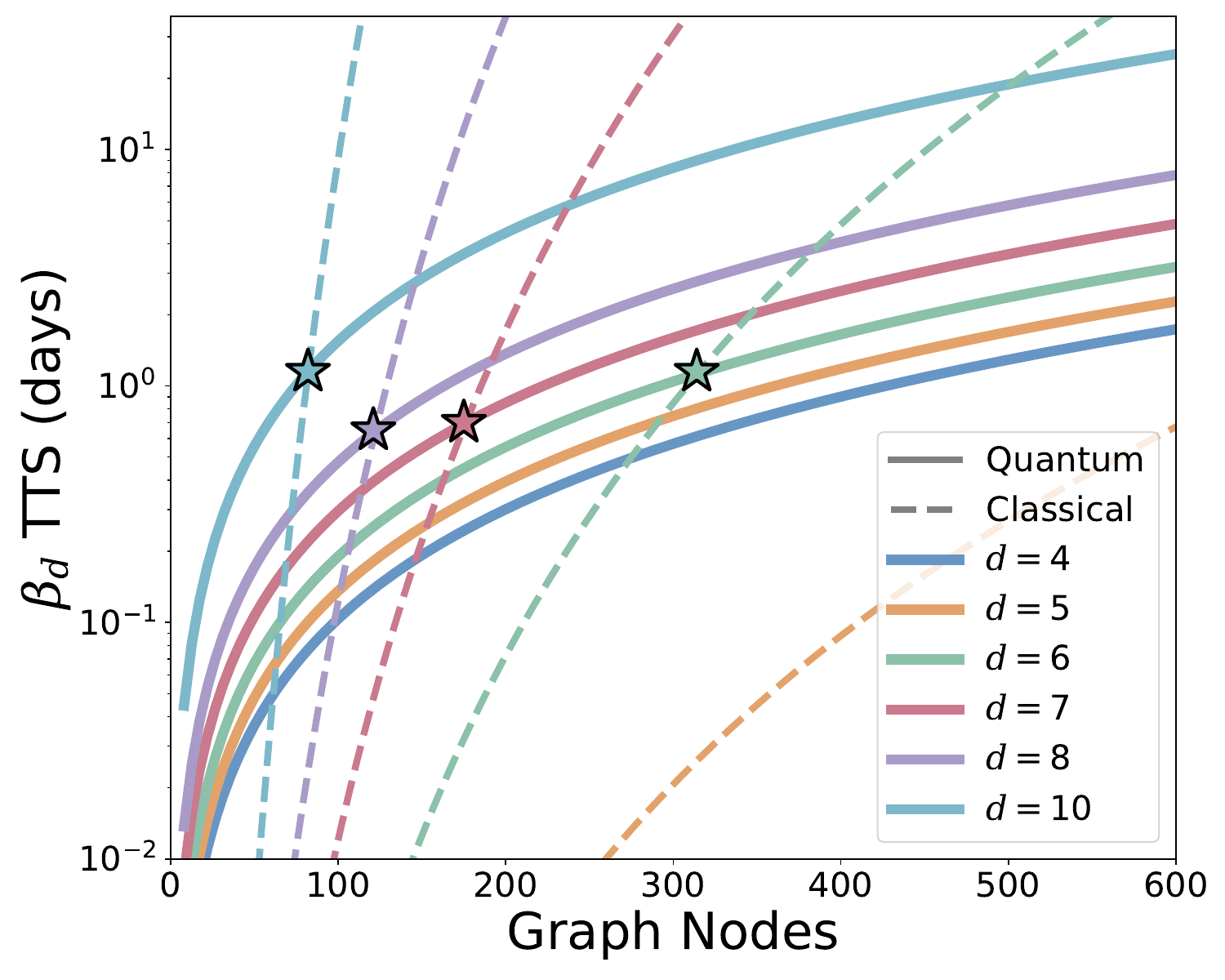}
    \caption{Crossover of quantum vs classical time-to-solution (TTS) for Betti numbers $\beta_d$ with $d=4-10$. We plot this for fixed edge density of $\zeta_2 = 0.90$ {\it(left)} and $\zeta_2=0.95$ {\it(right)}. }
    \label{fig:quantum_tts}
\end{figure}

In Fig.~\ref{fig:quantum_tts}, we compare the quantum and classical time-to-solution as a function of the number of graph nodes for two choices of edge density, $\zeta_2$, and for a range of homological degrees. We point out the point of the quantum-classical crossover time for both $\zeta_2=0.9$ and $\zeta_2=0.95$. Using our  assumptions, these crossover points occur between 70 and 1200 graph nodes and at times ranging between several hours and 10 days.

\begin{figure}[H]
    \centering
    \includegraphics[width=0.98\linewidth]{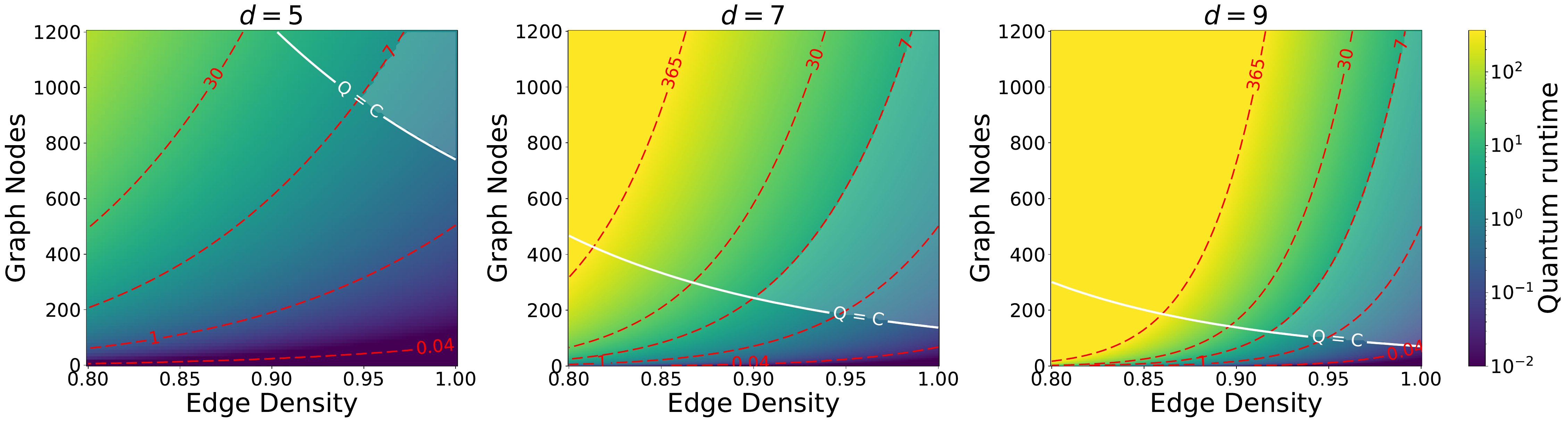}

    \caption{Heatmap of the time-to-solution as a function of both number of graph nodes, and edge density, of the quantum algorithm. This is shown for calculations of $\beta_d$ with $d = 5$, $7$ and $9$. The dashed red curves are fixed TTS contours of 1-day, 1 week, 1 month and 1 year. The solid white line shows the curve along which the classical TTS equals the quantum TTS. The quantum TTS strictly decreases as edge density increases, while the classical TTS is independent of edge density. The shaded region in each plot shows the regime where the quantum algorithm outperforms the classical with a total TTS $< 1$ week. }
    \label{fig:tts_comparison}
\end{figure}

In Fig.~\ref{fig:tts_comparison}, we show the time-to-solution of full phase-space parameterization of the problem. We plot the TTS as a function of edge density, and number of graph nodes for Betti numbers $\beta_d$ with $d=5,7,9$. The easiest path to quantum advantage is to push the edge density closest to $\zeta_2=1$, and choose the Betti dimension very high. However, to be of the most practical usefulness, we would like to push the edge density and Betti dimension as low as possible while still maintaining quantum advantage. We show the curve where the quantum and classical runtimes coincide for each plot.

\subsection{Simulation of Quantum TDA Circuits}

In this section, we implement the quantum TDA algorithm on the fMRI point clouds to demonstrate its accuracy and feasibility. 
We choose a point cloud randomly from the dataset, which belongs to one ROI of a patient. Based on the high correlation shown between Betti numbers and relative trace in Fig.~\ref{fig:hnorm_vs_RT_dim65}, we choose the graphs at $4$ different filtration values with $p \ge 0.6$. To make the simulation suitable for ideal quantum simulation on current simulators on a laptop, we compress the time series to $72$ dimensions and $12$ points. 

\begin{figure}[H]
    \centering
    {{\includegraphics[width=0.5\linewidth]{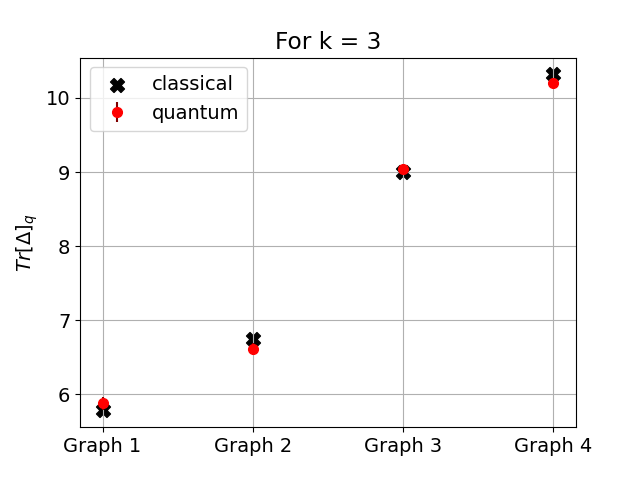}}~
    {\includegraphics[width=0.5\linewidth]{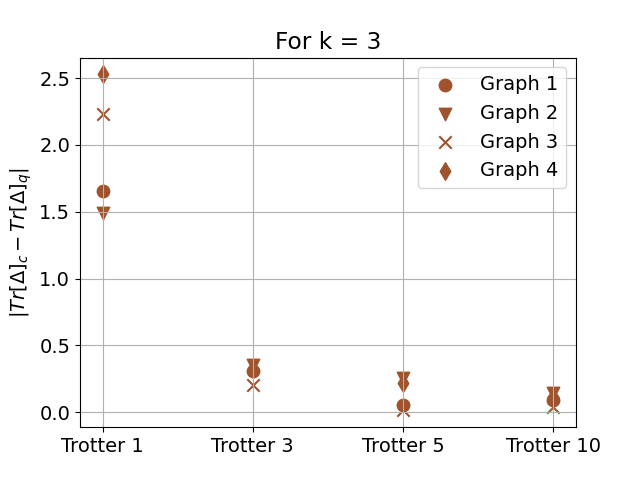}}}
    \caption{{\it (left)} Relative trace calculated for $4$ different graphs with $500$ initial vectors, $500$ shots, trotter level $=10$ and at $k=3$. The circuits have approximately $700 - 737$ CX gates. {\it (right)} Difference between the quantum output and the exact classical result at different Trotter levels for the $4$ graphs.}
    \label{fig:qTDA_simulation_fmri}
\end{figure}

This particular implementation of the algorithm has mid-circuit measurements, allowing us to perform the simulation with $16$ qubits (see Sec.~\ref{sec:qTDA_theory} for more details). With this circuit, a typical graph from our dataset requires $\sim 500$ CX gates for $k=3$ and trotter level $5$, depending on the number of edges in the graph. The more connected the graph is, the less deep is the circuit. We used trotter level=$10$ for the simulations in Fig.~\ref{fig:qTDA_simulation_fmri} with $500$ shots and $500$ random initial vectors.


Note that a higher trotter level gives more accurate results (closer to the exact output) but it also increases the circuit depth. One way to improve the accuracy of the results without changing the circuit depth is by reducing the parameter $t0$. However, this can also reduce the precision of the output. Increasing the number of initial vectors is another way to improve precision. Note that our quantum TDA outputs closely coincide with their corresponding exact values, highlighting the potential of the qTDA application for brain fMRI study, particularly in distinguishing graphs of high relative trace (lower Betti) from low relative trace (higher Betti).


\subsection{Trapped-Ion Hardware Demonstration}

Here, we report the result of running the QTDA Betti number estimation algorithm on IonQ's trapped ion quantum hardware. We performed these experiments on a Barium development system similar to the forthcoming Tempo-line systems, operating in a mode with 40 qubits. We ran the algorithm on two classes of simplicial complexes: (a) 8 node graphs with edge density $\zeta_2=0.57$, where we measured $\beta_1$ and (b) 16 node graphs with $\zeta_2 =0.70$ where we measured $\beta_3$. 

In order to run the algorithm on a 40 qubit QPU,  it is critical to use and the mid-circuit measurement (MCM) capability of this hardware. For example, the quantum circuits used to run the algorithm for the 16 node graphs used 32 qubits and used 4 rounds of mid-circuit measurements, where these MCM rounds measured 3,16,16 and 4 ancilla qubits respectively. Running this algorithm without the MCM capability would require a quantum circuit with at least 56 qubits. 

\begin{figure}[H]
    \centering
    \includegraphics[width=0.95\linewidth]{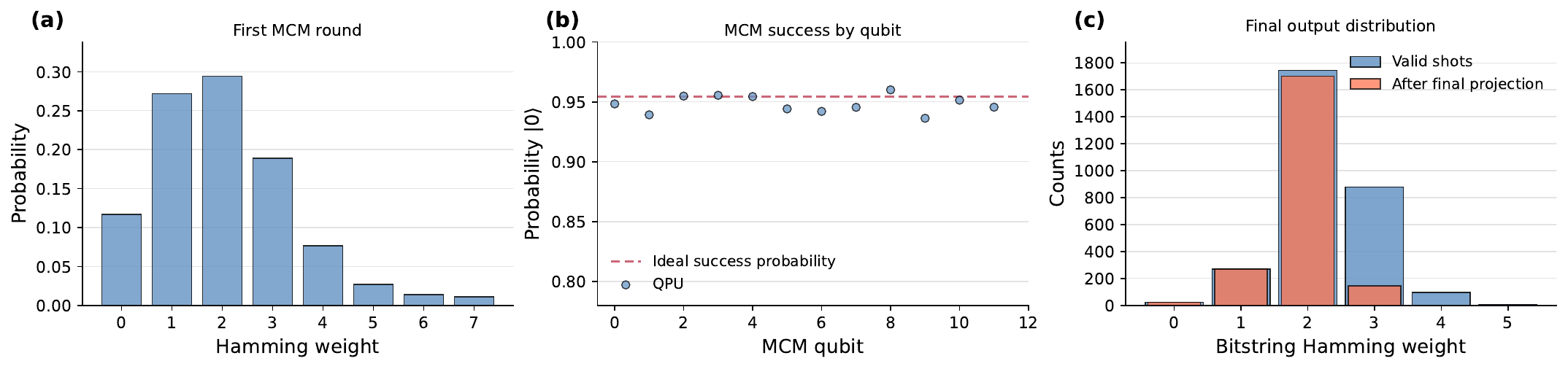}\\
    \includegraphics[width=0.95\linewidth]{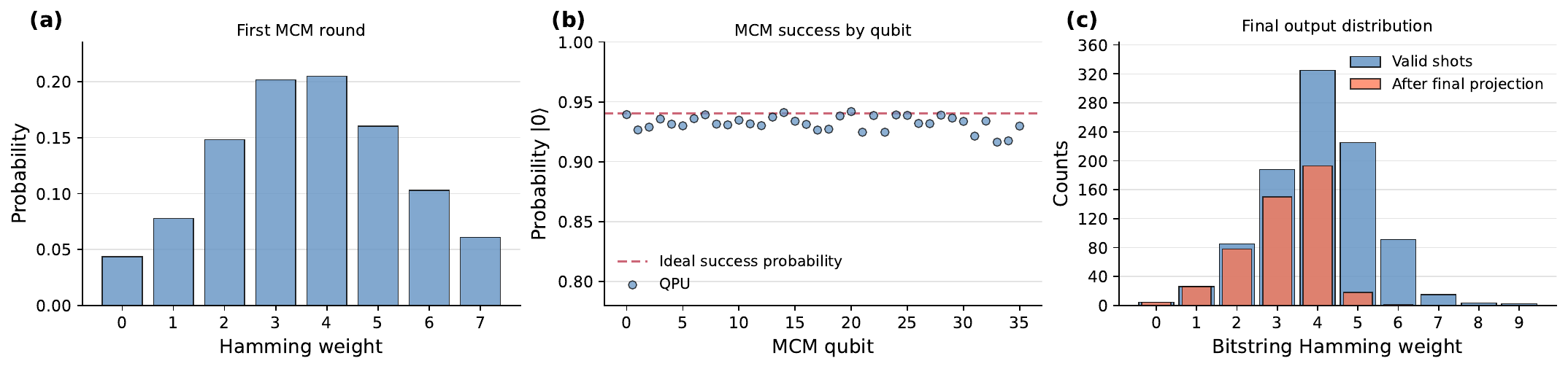}
    \caption{The mid-circuit and final measurement output of the QTDA algorithm for the $N=8$ {\it(top)} and $N=16$ {\it (bottom)} node graphs, corresponding to $16$ and $32$ qubits, respectively. The leftmost panel shows the output of the first MCM round where we perform QPE to measure the Hamming weight distribution and project into a fixed Hamming weight state. The middle panel shows the subsequent MCM round measurements on each qubit, where a measurement of the $\ket{0}$ state projects into the state representing our simplicial complex. Finally, the Hamming weight distribution of the end-of-circuit projective measurement on the graph qubits is shown. }
    \label{fig:MCM_hist}
\end{figure}

The output of the MCM rounds are shown in Fig.~\ref{fig:MCM_hist}. The first MCM round is implemented as part of the quantum phase estimation subroutine which counts the Hamming weight of the input quantum state and is used as part of the Dicke state preparation routine. We show the Hamming weight distribution of the initial product state used, which is prepared to be peaked around Hamming weight $k$ when running QTDA for $\beta_{k-1}$.  For both $k=2$ and $k=4$, we see that the measured distribution on the QPU is indeed peaked around the appropriate value. Post-selecting the quantum state on Hamming weight = $k$ allows us to prepare the appropriate Dicke state. 

The subsequent rounds of MCM are then used to project the Dicke state into the state representing our simplicial complex $\ket{\Gamma}$. As explained in Sec.~\ref{sec:resources}, this is performed using a rejection sampling routine, whereby we successfully project out all states not representing $k$-cliques of $\Gamma$ if we measure all $\ket{0}$ outputs on our ancilla qubits. For each ancilla qubit measured, we show the portion of the output where we measure $0$ on the QPU. The overall probability of success is given by the density of $k$-cliques in $\Gamma$, $\zeta_{k}$. Our measured success probability on the QPU is very close to the ideal rate $\zeta_{k}^{1/M}$, where $M$ is the number of measurements used. 

‹In Fig.~\ref{fig:MCM_hist} we also show the final end-of-circuit measurement distribution on the node qubits. This occurs after preparing the simplex state $\ket{\Gamma}$, and applying the Dirac operator $e^{iB\theta}$. Using the decomposition shown in Eq.~\ref{eq:eiB}, in the ideal case, the output distribution has support only on basis states with Hamming weight $k-1,k$ and $k+1$. Therefore, any bitstrings with Hamming weight outside these values were created due to hardware noise and can be discarded.  A final projection operator, $P_\Gamma$, is also be applied at the end of this circuit, which can be performed classically on the output histogram. In the ideal case, the Hamming weight $k$ and $k-1$ basis states in $\ket{\Psi} = e^{iB\theta}\ket{\Gamma}$ will be unaffected by this final projection, while the Hamming weight $k+1$ bitstrings will have support on bitstring both within and outside the $\Gamma$ space. Therefore, we can directly see the effect of hardware noise in the output histogram by comparing the population of states with forbidden Hamming weight, and comparing the real from expected number of shots that survive the final projection operator. We see that the circuits of the $k=2$, 8-node graphs are very close to the ideal distribution, indicating a very low level of hardware noise. From this output, we can see the effect of increased hardware noise as the circuit complexity increases in the $k=4$, 16-node graph circuits. In this case, the final output histogram contains a higher portion of shots in the $k=3$ and $k=4$ Hamming weight subspace which are not cliques of the input simplicial complex and therefore do not survive the final projection.

\begin{figure}[H]
    \centering
    \includegraphics[width=0.46\linewidth]{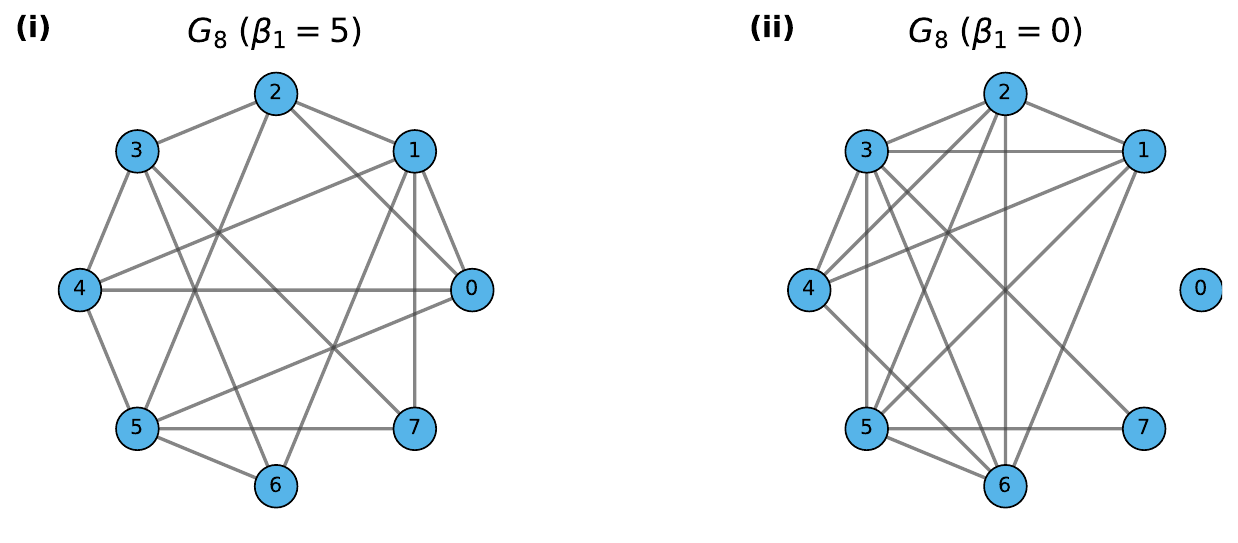}
    \quad 
    \vline
    \quad\,\,\,
    \includegraphics[width=0.44\linewidth]{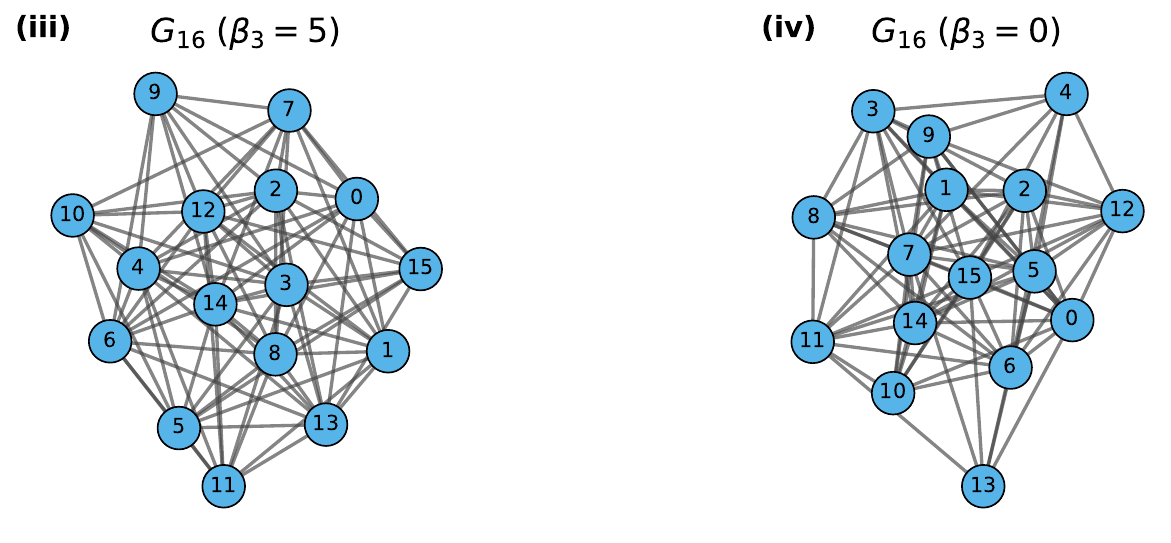}\\
    \vspace{0.1pc}
    \rule{16cm}{0.4pt}\\
    \includegraphics[width=0.98\linewidth]{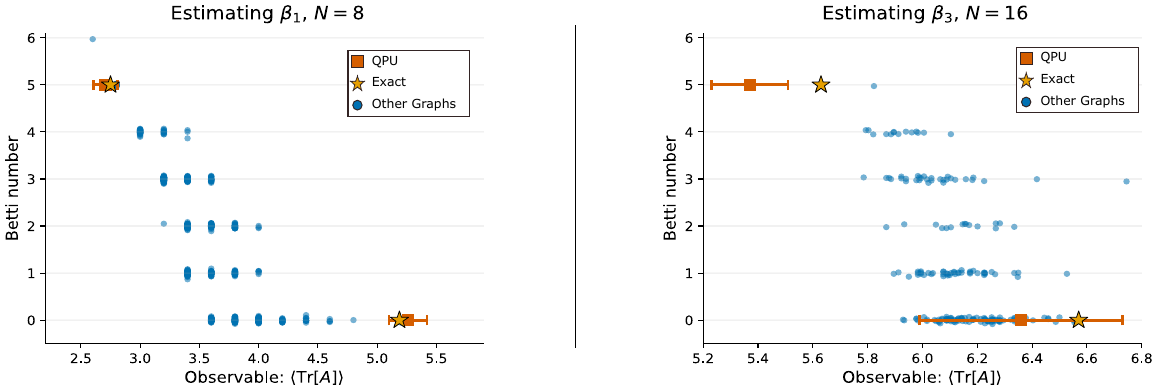}
    \caption{The distribution of the measured trace observable $\langle A \rangle = \operatorname{tr}[\Delta_k^\Gamma]$ vs the exact Betti number from the QPU experiments for graphs with $N=8$ and $N=16$ nodes, corresponding to $16$ and $32$ qubits, respectively. The quantum hardware runs were performed on two graphs for each case. We also show the exact value of the relative trace for a system of other randomly generated graphs with the same number of edges and approximately the same number of k-simplices. }
    \label{fig:qpu_result}
\end{figure}

We can process these histograms to calculate the final observable $A = \langle \text{tr}[\Delta^\Gamma_k]\rangle$. For each experiment, we consider a set of graphs with the physical parameters of number of nodes, N, edge density, $\zeta_2$, and k-clique density, $\zeta_k$, in a similar range. Based on the results discussed in Sec.~\ref{sec:rel_tr}, we can distinguish such graphs with different Betti number values based on the value of the observable $A=Tr[\Delta^\Gamma_k]$, the relative trace of the combinatorial Laplacian. In Fig.~\ref{fig:qpu_result}, we show that we are able to resolve the observable $A$ on the noisy QPU backend to sufficient accuracy to resolve the difference in Betti number between two graphs from this distribution. For both the $k=2$ and $k=4$ cases, we show the distribution of $\operatorname{tr}[\Delta_k]$ vs $\beta_{k-1}$ for the set of graphs with $N$ nodes and the the same value of $\zeta_2$ and $\zeta_k$. For each case, we ran the QTDA algorithm using graphs with relative trace values at the edges of the distribution. We find that in these two cases, we can resolve the values of $A$ with enough precision to distinguish the graphs based on their $\beta_{k-1}$ value. 

\section{Conclusions}

We explored the potential of quantum TDA for use in commercially relevant problems related to the analysis of time series data. We approached this problem from two perspectives: 1) determining if the information extracted by quantum TDA is useful, and 2) investigating if implementing a quantum algorithm for measuring this information is within reach. We provided evidence answering both of these questions in the affirmative, outlining a clear roadmap for achieving quantum advantage within this field. 

On the first question, we performed a classical TDA calculation to measure the high order homologies of time series data taken from financial markets and functional MRI scans. Our results provide evidence that there is important information in these high order homologies---up to $H_4$ was considered in our study---which can greatly improve the use of TDA as an early indicator for market crashes in the first case and as an informative feature for improved classification of neurodegenerative disorders. This expands upon earlier results which showed the usefulness of the $H_0$ and $H_1$ homologies for this task. Our work suggests that measuring even higher order homologies may further enhance the predictive power of this method. Classical algorithms for TDA become computational infeasible in these regimes, and so testing this hypothesis may require the application of quantum computing. 

On the second question, our analysis on the feasibility of the quantum algorithm suggests a clear pathway to implementing quantum circuits on near term devices which extract TDA features. We developed a new method for performing Betti number estimation, which is specifically tailored to reduce the circuit depths and shot counts required for this task, and proved this method out on a quantum computer similar to the forthcoming IonQ Tempo line. We showed that properties of simplicial complexes allow to measure the relevant operators within this framework, overcoming some of the obstacles which have plagued earlier versions of the quantum TDA algorithm. 

Together, these results paint a promising picture of how the quantum TDA algorithm may be used to gather new insights in data science, especially in the field of time series analysis. 

\section*{Acknowledgments}

We thank Shantanu Debnath, Mike Goldman, Ashay Patel, and Edwin Tham, and Ken Wright for assistance with quantum hardware executions and Dan Pompa for helpful discussions.
\clearpage
\bibliographystyle{unsrt}
\bibliography{references}
\clearpage

\end{document}